\documentclass[aps,prd,preprint,eqsecnum,
superscriptaddress,preprintnumbers,
showpacs,nofootinbib]{revtex4}
\usepackage{amsfonts,amssymb,amsmath,bm}
\usepackage{graphicx}

\newcommand{\be}{\begin{equation}}
\newcommand{\bea}{\begin{eqnarray}}
\newcommand{\ee}{\end{equation}}
\newcommand{\eea}{\end{eqnarray}}
\newcommand{\bpi}{\begin{picture}}
\newcommand{\bce}{\begin{center}}
\newcommand{\epi}{\end{picture}}
\newcommand{\ece}{\end{center}}

\def\chic#1{{\scriptscriptstyle #1}}

\def\gb{\bm{\Gamma}}

\begin{document}

\title{Non-perturbative comparison of QCD effective charges}

\author{A.~C. Aguilar}
\affiliation{Federal University of ABC, CCNH, 
Rua Santa Ad\'elia 166,  CEP 09210-170, Santo Andr\'e, Brazil.}
\author{D.~Binosi}
\affiliation{European Centre for Theoretical Studies in Nuclear
  Physics and Related Areas (ECT*), Villa Tambosi, Strada delle
  Tabarelle 286, I-38050 Villazzano (TN), Italy.}
\author{J. Papavassiliou}
\affiliation{Department of Theoretical Physics and IFIC, 
University of Valencia-CSIC,
E-46100, Valencia, Spain.}
\author{J.~Rodr\'iguez-Quintero}
\affiliation{Department of Applied Physics,
University of Huelva, E-21071 Huelva, Spain.}

\begin{abstract}

We  study the  non-perturbative behavior  of two  versions of  the QCD
effective  charge,  one  obtained   from  the  pinch  technique  gluon
self-energy,  and  one from  the  ghost-gluon  vertex.  Despite  their
distinct theoretical  origin, due  to a fundamental  identity relating
various of the ingredients  appearing in their respective definitions,
the two effective charges are  almost identical in the entire range of
physical  momenta, and coincide  exactly in  the deep  infrared, where
they  freeze at  a common  finite value.   Specifically,  the dressing
function of  the ghost propagator is  related to the  two form factors
in the Lorentz  decomposition of a certain Green's function,
appearing  in  a variety  of  field-theoretic  contexts.  The  central
identity, which is valid only in the Landau gauge, is derived from the
Schwinger-Dyson equations governing the dynamics of the aforementioned
quantities.   The   renormalization  procedure  that   preserves  the
validity  of  the  identity  is  carried  out,  and  various  relevant
kinematic limits  and physically motivated  approximations are studied
in  detail.  A  crucial ingredient  in this  analysis is  the infrared
finiteness of  the gluon  propagator, which is  inextricably connected
with  the  aforementioned freezing  of  the  effective charges.   Some
important issues related to the consistent definition of the effective
charge in  the presence of such  a gluon propagator  are resolved.  We
finally  present  a detailed  numerical  study  of  a special  set  of
Schwinger-Dyson    equations,    whose    solutions   determine    the
non-perturbative  dynamics   of  the  quantities   composing  the  two
effective charges.

\end{abstract}


\maketitle

\section{Introduction}

The infrared behavior of the QCD effective charge
is of considerable theoretical and phenomenological interest \cite{Cornwall:1982zr, Mattingly:1993ej, 
Cornwall:2009ud,Cornwall:1989gv}.
This quantity, when correctly defined, provides a continuous 
interpolation between two physically distinct regimes: the deep ultraviolet (UV), where 
perturbation theory works well, and the deep infrared (IR), where 
non-perturbative techniques must be employed. 
In fact, the effective charge is intimately connected with two 
phenomena that are of central importance to QCD: asymptotic freedom in the UV, 
and dynamical gluon mass generation in the IR~\cite{Cornwall:1982zr,Aguilar:2006gr}. 
Thus, while perturbatively it captures asymptotic freedom, 
it also exposes, due to the appearance of the Landau pole, the need 
of a non-perturbative regulating mechanism. 
Therefore,its low-energy behavior  
conveys valuable information about the way the theory cures  
the IR instabilities, namely  through the non-perturbative generation 
of a dynamical mass scale, which  tames the perturbative Landau pole.
What makes the effective charge such an interesting quantity to study is its 
strong dependence on the detailed characteristics of some of the most fundamental 
Green's functions of QCD, such as the gluon and ghost propagators.
Indeed, the basic ingredients that enter in its definition 
must contain the right information and be combined in a very precise way 
in order to endow the effective charge with the required physical and 
field-theoretic properties.

In this article we will focus on 
two characteristic definitions of the effective charge,  
frequently employed in the literature. 
The first definition is obtained within the 
pinch technique (PT) framework~\cite{Cornwall:1982zr,Cornwall:1989gv,Watson:1996fg}, and its correspondence~\cite{Denner:1994nn,Binosi:2002ft} with the 
background-field method (BFM)~\cite{Abbott:1980hw}. The PT effective charge, 
to be denoted by $\alpha_{\chic{\mathrm{PT}}}(q^2)$, constitutes the 
most direct non-abelian generalization of the familiar concept of the 
QED effective charge.   
The second definition of the QCD effective charge, to be denoted  by $\alpha_{\mathrm{gh}}(q^2)$, 
involves the ghost and gluon self-energies, in the Landau gauge, and in the kinematic 
configuration where the well-known Taylor non-renormalization theorem \cite{Taylor:1971ff, Marciano:1977su} becomes applicable.  
$\alpha_{\mathrm{gh}}(q^2)$  has  been  employed extensively 
in lattice studies~(see for instance
\cite{Bloch:2003sk, Boucaud:2008gn} and references therein), where  the Landau  gauge is the 
standard choice for   the  simulation   of  the   gluon   and  ghost
propagators, as well as in various investigations based on Schwinger-Dyson equations (SDEs)~\cite{von Smekal:1997vx, Boucaud:2006if}.
Even though the theoretical origin 
of the two aforementioned effective charges 
is rather distinct, it turns out that, quite remarkably, 
by virtue of a powerful non-perturbative identity, they 
are almost identical in the entire range of physical (euclidean) momenta.    
In fact, most interestingly, they are exactly equal in the deep IR 
({\it i.e.}, at vanishing momentum transfer, $q^2=0$). 

As we will see shortly, in the definition of the two effective charges appears 
a common ingredient, namely 
the gluon propagator (in the Landau  gauge), and two ingredients that 
are not common. These two non-common ingredients are,  
a-priori, not related  to each other;  the role of the aforementioned 
identity is to furnish a non-trivial connection between them.
Specifically, it relates 
the dressing function of the ghost propagator, denoted by $F(q^2)$,  
entering into the definition of $\alpha_{\mathrm{gh}}(q^2)$, with 
a certain function, denoted by $G(q^2)$, appearing 
in the definition of $\alpha_\chic{\mathrm{PT}}(q^2)$.  
The function $G(q^2)$ is the form-factor multiplying $g_{\mu\nu}$ 
in the Lorentz decomposition of a 
special Green's function, denoted by $\Lambda_{\mu\nu}(q)$, which 
appears in a variety of field-theoretic contexts. 
Most notably, $\Lambda_{\mu\nu}(q)$
enters in all ``background-quantum'' identities, i.e. the infinite tower of non-trivial relations connecting the 
BFM Green's functions to the conventional ones~\cite{Grassi:1999tp,Binosi:2002ez}. 
Notice also that $G(q^2)$ 
plays a central role in the new SDEs derived 
within the PT framework~\cite{Binosi:2007pi}; due to the special properties of 
the Green's functions involved, these new SDEs can be truncated in a manifestly gauge 
invariant way~\cite{Aguilar:2006gr}. The identity in question 
connects the two non-common ingredients of the two charges, $F(q^2)$ and $G(q^2)$, to the second 
form factor of $\Lambda_{\mu\nu}(q)$, denoted by $L(q^2)$, in the way shown in Eq.~(\ref{funrel}).

To the best of our knowledge, the identity of Eq.~(\ref{funrel}) was first derived in~\cite{Kugo:1995km}, in connection with the so-called Kugo-Ojima confinement criterion~\cite{Kugo:1979gm}. 
The same identity was proved in~\cite{Grassi:2004yq}, where the general algebraic properties of $SU(N)$ Yang Mills theories in the background Landau gauge 
were studied; however, no connection with the conventional $R_\xi$ Landau gauge was established. 
More recently, it was revisited in~\cite{Kondo:2009ug},  where a new relation between the Kugo-Ojima parameter and the Gribov-Zwanziger horizon function has been advocated.
However, to date, the dynamical equations for the quantities appearing in this identity
remain largely unknown.

In the present  work we derive the central  identity starting from the
SDEs  that  govern the  dynamics  of  the  relevant functions,  namely
$F(q^2)$,  $G(q^2)$, and  $L(q^2)$.  These  SDEs allow  for  a detailed
study  of the  individual properties  of these  three  functions, both
perturbatively  and  non-perturbatively.  Most importantly,  they
expose the  way these  functions depend on  the gluon  propagator, and
furnish  a  self-consistent framework  for  studying  how an  IR
finite  gluon  propagator affects  their  IR properties.   These
properties, in turn, are  responsible for the mild discrepancy between
the two effective charges mentioned above.

The paper  is organized as  follows.  In Section~\ref{seffch}, 
after introducing the necessary notation and definitions,   
we outline the basic theoretical ingredients entering into 
the construction of the two 
(dimensionful) renormalization-group (RG) invariant quantities, from which the   
two (dimensionless) effective charges, 
$\alpha_\chic{\mathrm{PT}}(q^2)$  and $\alpha_\mathrm{gh}(q^2)$ will be extracted.
 Then, we focus on the timely question of how to identify 
the correct non-perturbative  scale in the
presence    of    an   IR    finite    gluon   propagator.    
The central identity  of the paper is derived in
Section~\ref{giSDE}, starting from the defining SDEs.  
The renormalization    procedure that preserves the validity of the 
identity is carried out, and various properties are studied in the 
UV and IR kinematic limits; most notably, 
we establish that if the gluon   propagator is IR finite, then 
$L(0)=0$. The implications of the identity 
on the two  effective charges are discussed, and a 
relation between them is established, which is valid for the 
entire range of euclidean momenta.   
A detailed numerical analysis and
comparison of  the two effective charges  at different renormalization
scales is  carried out in~Section~\ref{numan},  using as an  input the
non-perturbative solutions  of the  SDEs corresponding to  the various
functions    appearing    in    their    definition.    Finally,    in
Section~\ref{concl} we present our conclusions.

\section{\label{seffch} Two non-perturbative effective charges}

In this section we will first  introduce some of the basic filed-theoretic 
ingredients necessary for the definition of the two effective charges we want to study.
Then, we will briefly outline the basic construction and the assumptions involved 
in the definition of either charge. Finally, we will discuss in detail 
the important issue of how to extend the two definitions to the non-perturbative regime, and, 
in particular, the identification of the correct scale in the presence of an 
IR-finite gluon propagator.   

\begin{figure}[!t]
\includegraphics[width=10cm]{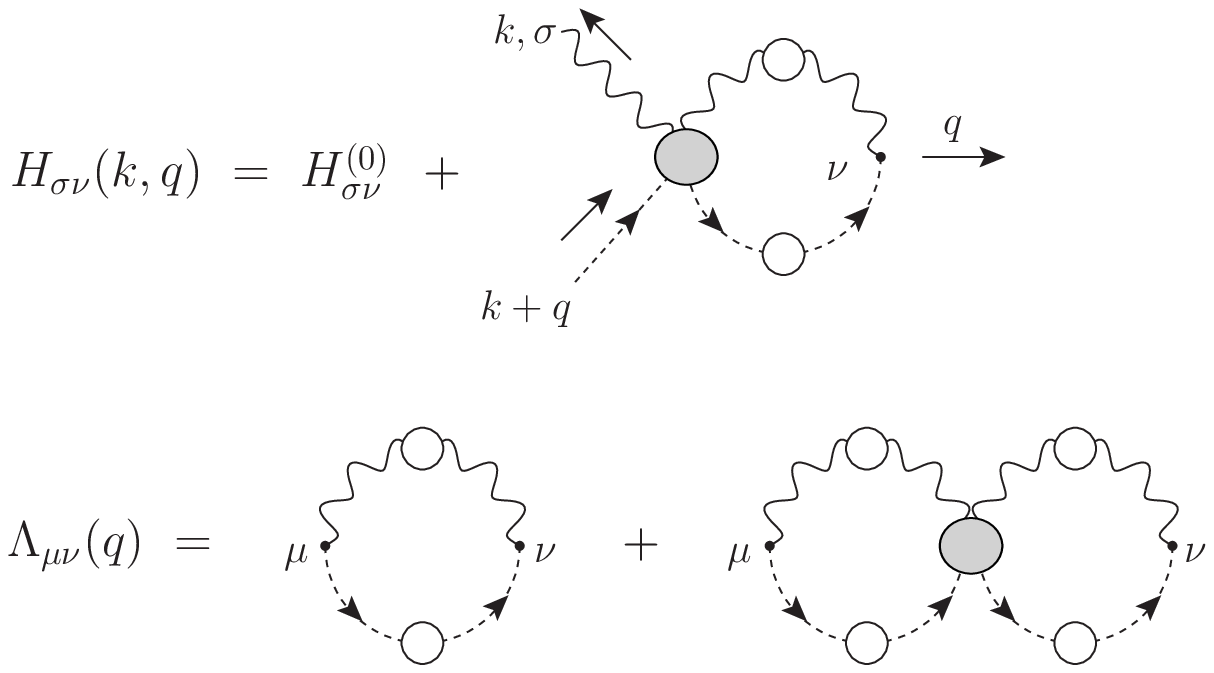}
\caption{Diagrammatic representation of the functions $H$ and $\Lambda$.}
\label{fig:Lambda_aux}
\end{figure}

\subsection{Definitions and ingredients}

Let  us   first introduce  the notation and define some of the basic quantities entering into 
the problem under study. 

In the covariant  renormalizable ($R_\xi$) gauges, the gluon
propagator $\Delta_{\mu\nu}(q)$  has the form  
\be \Delta_{\mu\nu}(q)=-i\left[ P_{\mu\nu}(q)\Delta(q^2) +\xi\frac{q_\mu q_\nu}{q^4}\right],
\label{prop_cov}
\ee
where $\xi$ denotes the gauge-fixing parameter, and 
\mbox{$P_{\mu\nu}(q)= g_{\mu\nu} - q_\mu q_\nu /q^2$}
is the usual transverse projector. Finally, $\Delta^{-1}(q^2) = q^2 + i \Pi(q^2)$, 
with  $\Pi_{\mu\nu}(q)=P_{\mu\nu}(q)\Pi(q^2)$ the gluon self-energy  (notice that since $\Pi(q^2)$ has been defined with the imaginary factor $i$ pulled out in front, it is simply given by the corresponding Feynman diagrams in Minkowski space). 
In addition, the full ghost propagator $D(q^2)$ and its dressing function $F(q^2)$
are related by 
\be
D(q^2)= \frac{iF(q^2)}{q^2}.
\label{ghostdress}
\ee
Moreover, the all-order ghost vertex (after factoring out the color structure and the coupling constant $g$) 
will be denoted by  $\gb_\mu(k,q)$ with $k$ representing the momentum of the gluon and $q$ the one of the anti-ghost. 
The tensorial structure is given by
\be
-{\gb}_{\mu}(k,q) =  B_1(k,q) q_{\mu} + B_2(k,q) k_{\mu}.
\label{Gtens}
\ee 
Thus, at tree-level $\gb^{(0)}_\mu(k,q)=\Gamma_\mu(k,q)=-q_\mu$.

An important ingredient for what follows is 
the two-point function $\Lambda_{\mu\nu}(q)$ represented in Fig.~\ref{fig:Lambda_aux}, defined by
\bea
 \Lambda_{\mu \nu}(q) &=&-i g^2C_A
\int_k H^{(0)}_{\mu\rho}
D(k+q)\Delta^{\rho\sigma}(k)\, H_{\sigma\nu}(k,q),
\nonumber \\
&=& g_{\mu\nu} G(q^2) + \frac{q_{\mu}q_{\nu}}{q^2} L(q^2),
\label{LDec}
\eea
where $C_{\rm {A}}$ the Casimir eigenvalue of the adjoint representation
[$C_{\rm {A}}=N$ for $SU(N)$], 
and \mbox{$\int_{k}\equiv\mu^{2\varepsilon}(2\pi)^{-d}\int\!d^d k$}, 
with $d=4-\epsilon$ the dimension of space-time. 
The scalar function $G(q^2)$ appearing in the equation above allows the 
connection between the conventional and BFM-PT gluon propagators, 
and is known to play a central role in the PT formulation of the SDE.

The function $H_{\mu\nu}(k,q)$ (see  Fig.~\ref{fig:Lambda_aux} for a diagrammatic definition)
is in fact a familiar object~\cite{Marciano:1977su}: 
it appears in the all-order Slavnov-Taylor identity
satisfied by the standard  three-gluon vertex, and
is related to the full gluon-ghost vertex by 
\be
q^\nu H_{\mu\nu}(k,q)=-i\gb_{\mu}(k,q).
\label{qH}
\ee
At tree-level, $H_{\mu\nu}^{(0)} = ig_{\mu\nu}$.
Finally, using the most general Lorentz decomposition of  $H_{\mu\nu}$, 
\be
-iH_{\mu\nu}(k,q) =  A_1(k,q) g_{\mu\nu}  + A_2(k,q) q_{\mu}q_{\nu} + A_3(k,q) k_{\mu}k_{\nu} 
+ A_4(k,q) q_{\mu}k_{\nu} +  A_5(k,q) k_{\mu}q_{\nu},
\label{Htens}
\ee
we obtain from (\ref{Gtens}) and (\ref{qH}) two constrains for the various form-factors, namely  
\bea
B_1(k,q) &=& A_1(k,q) + q^2 A_2(k,q) + (k \cdot q) A_4(k,q),
\nonumber\\
 B_2(k,q)&=& (k \cdot q) A_3(k,q) + q^2 A_5(k,q).
\label{con1}
\eea

\subsection{The pinch technique effective charge}

The QCD effective charges constructed within the PT 
uses QED as the basic reference point~\cite{Itzykson:rh}. 
In QED, one begins by considering  
the unrenormalized photon self-energy   $\Pi^{0}_{\alpha\beta}(q) =  q^2
P_{\alpha\beta}(q)\Pi^{0}(q^2)$,  where
$P_{\alpha\beta}(q)=g_{\alpha\beta}-q_\alpha q_\beta/q^2$ and  
$\Pi^{0} (q^2)$ is  a
gauge-independent function to all  orders in perturbation theory.  After
Dyson   summation,  we obtain   the (process independent)  dressed
photon propagator between conserved external currents
$\Delta^{0}_{\alpha\beta}(q)\ = 
(g_{\alpha\beta}/q^2)\Delta^{0}(q^2)$, with 
$\Delta^{0}(q^2)= -i[1+i\Pi^{0}(q^2)]^{-1}$.
The renormalization procedure introduces 
the standard relations  between  renormalized    and unrenormalized
parameters:  
$e = Z_{e}^{-1} e_{0} = Z_f Z_A^{1/2} Z_1^{-1} e_{0}$
and 
$1+i\Pi (q^2)= Z_A [1+i\Pi ^{0} (q^2)]$, 
where $Z_A$ ($Z_f$) is the wave-function renormalization 
constants of the
photon (fermion),  $Z_1$ the vertex renormalization, 
and $Z_{e}$ is the charge renormalization constant. 
The  Abelian gauge
symmetry  of the   theory   gives  rise   to  the    fundamental 
Ward identity (WI) 
$q^{\alpha}\Gamma^{0}_{\alpha}(p,p+q)= S_{0}^{-1}(p+q)-S_{\rm
o}^{-1}(p)$, where 
$\Gamma^{0}_{\alpha}$ and  $S_{0} (k)$ are the  unrenormalized
all orders    
photon-electron vertex and electron propagator, respectively.
The requirement  that     the   renormalized vertex       $\Gamma_{\alpha} 
=
Z_{1}\Gamma^{0}_{\alpha}$   and  the  renormalized  self-energy   $S  =
Z_{f}^{-1} S_{0}$ satisfy the same identity, implies 
$Z_{1}=Z_{f}$, from which immediately follows that 
$Z_{e}\ =\ Z_{A}^{-1/2}$. Given these relations between 
the renormalization constants, and after pulling out the trivial
factor $g_{\alpha\beta}/q^2$, 
we can form the renormalization group invariant combination, known as
the effective charge, 
\be
\alpha(q^2)=
\frac{e_0^2}{4\pi}\Delta^{0}(q^2)=
\frac{e^2}{4\pi} \Delta(q^2).  
\label{alphaqed}
\ee

In QCD, the crucial equality $Z_1=Z_f$ does not
hold, because the WIs are replaced by 
the more complicated Slavnov-Taylor identities (STIs),  
involving ghost Green's functions \cite{Marciano:1977su, Itzykson:rh}.  
Furthermore, the gluon self-energy  
depends on  the gauge-fixing parameter, already  at one-loop
order. These facts render the QCD generalization of a QED-like 
effective charge more complicated; however, the theoretical framework of the
PT makes this  definition possible~\cite{Cornwall:1982zr,Watson:1996fg}.
The PT rearranges the conventional gauge
dependent $n$-point Green's functions, to construct individually gauge
independent Green's functions, which, in addition, obey naive (ghost free) WIs .
One important point, explained in detail in the literature, is the (all-order) correspondence 
between the PT and the Feynman gauge of the BFM~\cite{Denner:1994nn, Binosi:2002ft}. 
In fact, using the 
methodology introduced in ~\cite{Pilaftsis:1996fh}, one can generalize the PT construction
in such a way as to reach diagrammatically 
any value of the gauge fixing parameter of the BFM, and in particular the Landau gauge.
In what follows we employ the aforementioned generalization of the PT, 
given that the identity we will eventually derive is valid only in the Landau gauge.

\begin{figure}[!t]
\includegraphics[width=14cm]{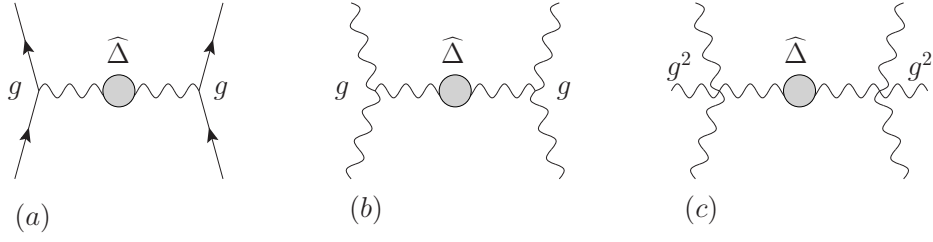}
\caption{The universal PT coupling.}
\label{fig:pt_coup}
\end{figure}

The PT definition of the effective charge 
relies on the construction of an universal ({\it i.e.}, process-independent) 
effective gluon propagator,  
which captures the running of the QCD $\beta$ function, exactly as happens with the 
vacuum polarization in the case of QED  (See Fig.~\ref{fig:pt_coup}).
To fix the ideas,  the PT one-loop gluon self-energy reads
\be 
\widehat\Delta^{-1}(q^2)= q^2\left[1+ b g^2\ln\left(\frac{q^2}{\mu^2}\right)\right],
\label{rightRG}
\ee
where  $b = 11 C_A/48\pi^2$  is the first coefficient of the QCD $\beta$-function. 
Due to the Abelian WIs satisfied by the PT effective Green's functions, the 
 new propagator-like quantity $\widehat\Delta^{-1}(q^2)$ absorbs all  
the RG-logs, exactly as happens in QED with the photon self-energy.
Then, the renormalization 
constants of the gauge-coupling and of the PT gluon self-energy, 
defined as 
\bea
g(\mu^2) &=&Z_g^{-1}(\mu^2) g_0 ,\nonumber \\
\widehat\Delta(q^2,\mu^2) & = & \widehat{Z}^{-1}_A(\mu^2)\widehat{\Delta}_0(q^2), 
\label{conrendef}
\eea
where the ``0'' subscript indicates bare quantities, satisfy the 
QED-like relation 
\be
{Z}_{g} = {\widehat Z}^{-1/2}_{A}.
\label{ptwi}
\ee
Of course, $Z_g$ must be obtained under a given 
renormalization prescription, and the PT gluon self-energy 
will be then renormalized imposing~(\ref{ptwi}). 
Thus, regardless of  the renormalization prescription chosen, 
the product 
\be
{\widehat d}_0(q^2) = g^2_0 \widehat\Delta_0(q^2) = g^2(\mu^2) \widehat\Delta(q^2,\mu^2) = {\widehat d}(q^2), 
\label{ptrgi}
\ee
retains the same form before and after renormalization, {\it i.e.}, it 
forms a RG-invariant ($\mu$-independent) quantity~\cite{Cornwall:1982zr}.

For asymptotically large momenta one may extract from ${\widehat d}(q^2)$
a dimensionless quantity by writing,
\be
{\widehat d}(q^2) = \frac{\overline{g}^2(q^2)}{q^2},
\label{ddef1}
\ee
where $\overline{g}^2(q^2)$ is the RG-invariant effective charge of QCD; at one-loop
\be
\overline{g}^2(q^2) = \frac{g^2}{1+  b g^2\ln\left(q^2/\mu^2\right)}
= \frac{1}{b\ln\left(q^2/\Lambda^2_\chic{\mathrm{QCD}}\right)}.
\label{effch}
\ee
where $\Lambda_\chic{\mathrm{QCD}}$ denotes an RG-invariant mass scale of a few hundred ${\rm MeV}$.

Eq.~(\ref{ptrgi}) is a non-perturbative relation; therefore it can serve unaltered as the starting 
point for extracting a non-perturbative effective charge, 
provided that one has information on the IR behavior of the PT-BFM gluon propagator $\widehat\Delta(q^2)$.
Interestingly enough, non-perturbative information on the {\it conventional} gluon propagator $\Delta(q^2)$ 
may also be used, by virtue of a general relation connecting $\Delta(q^2)$ and $\widehat\Delta(q^2)$.
Specifically, a formal all-order relation known as ``background-quantum'' identity ~\cite{Grassi:1999tp, Binosi:2002ez}
states that 
\be
\Delta(q^2) = 
\left[1+G(q^2)\right]^2 \widehat{\Delta}(q^2).
\label{bqi2}
\ee
Note that, due to its BRST origin, the above relation must be preserved after renormalization. 
Specifically, denoting by $Z_\Lambda$ the (yet unspecified) renormalization constant relating 
the bare and renormalized functions, $\Lambda_0^{\mu\nu}$ and $\Lambda^{\mu\nu}$, through
\be
\Lambda^{\mu\nu}(q,\mu^2)=Z_\Lambda(\mu^2) \Lambda_0^{\mu\nu}(q),
\label{Lamrel}
\ee
then from Eqs.~(\ref{bqi2}) and (\ref{ptwi}) follows the additional relation 
\be
Z_g^{-1} = Z_A^{1/2} Z_\Lambda ,
\label{extrel}
\ee 
which is useful for the comparison with the coupling discussed 
in the following subsection.

It is now easy to verify, at lowest order, that 
the $1+G(q^2)$ obtained from Eq.~(\ref{LDec})  restores the $\beta$ function coefficient  
in front of UV logarithm. In that limit~\cite{Aguilar:2008xm}
\bea
1+G(q^2) &=& 1 +\frac{9}{4}
\frac{C_{\rm {A}}g^2}{48\pi^2}\ln\left(\frac{q^2}{\mu^2}\right),\nonumber \\
\Delta^{-1}(q^2) &=& q^2 \left[1+\frac{13}{2}
\frac{C_{\rm {A}}g^2}{48\pi^2}\ln\left(\frac{q^2}{\mu^2}\right)\right].
\label{pert_gluon}
\eea
Using  Eq.~(\ref{bqi2}) we therefore recover the $\widehat{\Delta}^{-1}(q^2)$ 
of Eq.~(\ref{rightRG}), as we should. 

Then, non-perturbatively, one substitutes into  Eq.~(\ref{bqi2}) the $ 1+G(q^2)$ and $\Delta(q^2)$ 
obtained from either the lattice or SD analysis, to obtain $\widehat{\Delta}(q^2)$.  
This latter quantity is the non-perturbative generalization of Eq.~(\ref{rightRG}); 
for the same reasons explained above, the combination 
\be
\widehat{d}(q^2)= \frac{g^2 \Delta(q^2)}{\left[1+G(q^2)\right]^2},
\label{rgi}
\ee
is an RG-invariant quantity.

\subsection{The effective charge from the ghost-gluon vertex}

In the previous subsection it has become clear that 
the PT construction involves a particular combination 
of two point functions only, with no explicit reference to any of the 
full vertices of the theory.
Thus, as happens in QED, 
the  effective charge so obtained is universal ({\it i.e.}, it does not depend 
on the details of the process where the PT propagator is embedded), and  
depends naturally on a single scale, namely the physical momentum exchange 
of a given process. 

In principle, a definition for the QCD effective charge can be obtained starting 
from the various QCD vertices\footnote{In fact, as has been explained in detail in~\cite{Binger:2006sj}, an effective charge may also be defined from the gauge-invariant three-gluon vertex~\cite{Cornwall:1989gv}.}, {\it i.e.}, 
the ghost-gluon vertex, the three- and the four-gluon vertices, the quark-gluon vertex, etc
\cite{Alkofer:2004it}. However, a priori, such a construction 
involves more than one scales, and further assumptions about their values need be introduced, in order 
to express the charge as a function of a single variable. 
As a general rule in all such a constructions one 
identifies a RG-invariant quantity formed by a judicious combination of the 
vertex form-factor and the  self-energies associated  with the fields entering into the vertex. 
 Let us assume, for example, a vertex with three fields, $\Phi_{i}(q_i)$, $i=1,2,3$,   
entering ( with \mbox{$q_1+q_2+q_3=0$}). 
Denoting the corresponding propagators 
by $\Delta_i(q_i)$, the relevant vertex form-factor by $V(q_1,q_2,q_3)$,  
by $Z_{i}$ the corresponding wave-function renormalization constants,   
and by $Z_V$ the vertex renormalization constant, one 
can renormalize the coupling such that\footnote{In the MOM prescription, for instance, $Z_g$ is determined by requiring that the renormalized vertex at the subtraction point assumes its tree-level value.} \mbox{$Z_g=Z_V (Z_1 Z_2 Z_3)^{-1/2}$}, from which follows that the 
combination 
\be
\widehat{r}(q_1,q_2,q_3) \equiv g^2  V^2(q_1,q_2,q_3)
 \Delta_1(q_1)\Delta_2(q_2) \Delta_3(q_3) \,,
\label{rggeneric}
\ee
is a RG-invariant quantity. 
As mentioned above, the complication with this definition is that $\widehat{r}(q_1,q_2,q_3)$
is a function of two kinematic variables. 
Thus, some additional assumption on the preferred 
kinematic configuration is usually introduced, such as, 
for example, 
\mbox{$q_1^2=q_2^2=q_3^2=q^2$} 
(and therefore \mbox{$q_1 \cdot q_2 = q_1 \cdot q_3 = q_2 \cdot q_3 = - q^2/2$)}, which  fully specifies the kinematic of the renormalization point.

For the case of the ghost-gluon vertex, let us define in general 
the following renormalization constants
\bea
\Delta(q^2,\mu^2)&=& Z^{-1}_{A}(\mu^2)\Delta_0(q^2),\nonumber\\
F(q^2,\mu^2)&=& Z^{-1}_{c}(\mu^2)F_0(q^2),\nonumber\\ 
\gb^\nu(k,q,\mu^2)&=& Z_1(\mu^2)\gb^\nu_0(k,q),\nonumber\\ 
g_0&=& Z_{g^{\prime}}(\mu^2)g^{\prime}.
\label{renconst}
\eea
Notice that a priori $Z_{g^{\prime}}$ defined as
$Z_{g'}= Z_1 Z_A^{-1/2} Z_c^{-1}$,
does not have to coincide with the 
$Z_{g}$ introduced in (\ref{conrendef}); however, as we will see 
in the next section, they do coincide by virtue of the basic identity 
we will derive there.

\begin{figure}[!t]
\includegraphics[width=11cm]{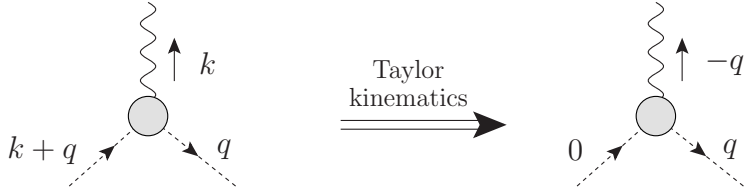}
\caption{The ghost-gluon vertex and the Taylor kinematics.}
\label{fig:ghostv}
\end{figure}

In the Landau gauge, the form factor
$B_1$ of Eq.~(\ref{Gtens}) is UV finite at one-loop, 
and therefore, no infinite renormalization constant needs to be introduced at that order;
of course, $B_2$ must be UV finite in all gauges, and to all orders, otherwise the
theory would be non-renormalizable. 
In order to obtain information about the UV behavior of $B_1$ beyond one-loop, 
one usually invokes the non-renormalization theorem of Taylor, which states 
that for vanishing ghost momentum (see  Fig.~\ref{fig:ghostv}), one has that 
\mbox{$B_1(-q,q)+ B_2(-q,q) =1$}, to all orders in perturbation theory.
Given that $B_2$ is finite to all orders (for any kinematic configuration), 
it follows that $B_1(-q,q)$ is also {\it finite} to all orders. 

In particular, for the Taylor (vanishing incoming ghost momentum) 
kinematics, $Z_1$ will be determined as above explained by demanding 
that the relevant form factor be equal to its tree-level value after 
renormalization\footnote{
Recall that the form factor emerging at the Taylor kinematic limit 
$k_\mu \to -q_\mu$ is $B_1 + B_2$.}, {\it i.e.}, $Z_1 \left[ (B_1(-q,q) + B_2(-q,q) \right] = 1$. Then, one will have that 
\be
Z_1= Z_{g^{\prime}} Z_A^{1/2}Z_c=1,
\label{Z1T}
\ee
from which follows that 
\be
Z_{g^{\prime}}^{-1} = Z_A^{1/2}Z_c. 
\label{mores}
\ee
Thus, the product
\be
\widehat{r}(q^2) \ = \ {g'}^2 \Delta(q^2;\mu^2) F^2(q^2;\mu^2) 
\ = \ g_0 \Delta_0(q^2) F^2_0(q^2),
\label{rg2}
\ee
forms either a dimensionful $\mu$-independent combination 
or a UV cut-off independent one. Provided that we renormalize the 
propagators in the MOM scheme with Taylor kinematics (named as ``Taylor scheme'' 
in \cite{Boucaud:2008gn}), $\widehat{r}(q^2)$ is a RG-invariant combination.

Therefore, for asymptotically large $q^2$, in analogy to Eq.~(\ref{ddef1}) one can define  
an alternative QCD running coupling as 
\be
\widehat{r}(q^2)=\frac{\overline{g}_{\mathrm{gh}}^2(q^2)}{q^2}.
\ee 
Notice that $\overline{g}_{gh}(q^2)$ has been shown to display the same behavior 
at any loop order as the ghost-gluon coupling for the Taylor kinematics 
(see Fig.~\ref{fig:ghostv}) in~\cite{Boucaud:2008gn}.

Using then Eq.~(\ref{pert_gluon}), and the fact that
\be
D^{-1}(q^2) = q^2 \left[1+\frac94\frac{C_{\rm {A}}g^2}{48\pi}\ln\left(\frac{q^2}{\mu^2}\right)\right],
\ee
it is straightforward to verify that  $\overline{g}_\mathrm{gh}(q^2)$ and  $\overline{g}(q^2)$ displays the same one-loop 
behavior, since, perturbatively the function $1+G(q^2)$ is the inverse of the ghost dressing function $F(q^2)$.
As we will see in the next section, this is nothing more than the one-loop manifestation 
of the more general identity relating $G(q^2)$ and $F(q^2)$.

\subsection{Effective charges from massive gluon propagators}

It is clear from the above analysis that before actually defining the effective 
charge with either method one 
constructs two {\it dimensionful} RG-invariant quantities, given in Eq.~(\ref{rgi}) and Eq.~(\ref{rg2}), 
with  mass dimension -2.
These two quantities share an important common ingredient, namely the 
scalar cofactor of the gluon propagator, $\Delta(q^2)$, 
which actually sets the scale. 
The next step is to extract a {\it dimensionless} quantity, that would correspond to the non-perturbative 
effective charge. Perturbatively, {\it i.e.}, for asymptotically large momenta, 
it is clear that the mass scale is saturated simply by $q^2$, the bare gluon propagator,  
and the effective charge is defined by pulling a $q^{-2}$ out of the corresponding  RG-invariant 
quantity\footnote{This is equivalent to the standard MOM prescription for the coupling definition.}.

Of course, as has been firmly established by now, 
in the IR the gluon propagator becomes effectively massive; 
therefore, particular care is needed in deciding exactly what 
combination of mass-scales ought to be pulled out.
The correct procedure in such a case  
has been explained long time ago in the pioneering  
work of Cornwall~\cite{Cornwall:1982zr}, and has been applied in various occasions~\cite{Aguilar:2007ku}: 
a ``massive'' propagator, of the form  $[q^2 + m^2(q^2)]^{-1}$ 
must be pulled out, where $m^2(q^2)$ is a dynamical ({\it i.e.}, momentum-dependent) 
mass\footnote{
Within the MOM philosophy one may implement 
the correct prescription by imposing 
\mbox{$\Delta^{-1}(\mu^2)=\mu^2+m^2(\mu^2)$}
as the (non-perturbative) MOM renormalization condition for the gluon 
propagator. This prescription is equivalent to 
the standard one in the UV,  while in the IR it introduces to 
the anomalous dimensions  genuine 
non-perturbative (Borel non-analytical) terms of the type
$\exp{(-1/g_R(q^2))}$,  which vanish as 
$q^2 \to \infty$.}.

Before applying this (correct) prescription to the two RG-invariant quantities in question, 
it is interesting to compare the situation with the 
more familiar, and conceptually more straightforward, case of the 
electroweak sector, where the corresponding gauge bosons ($W$ and $Z$) are also massive, 
albeit it through an entirely different mass generation mechanism. 
Specifically, while the $W$ and $Z$ bosons become massive at tree-level, through the 
standard Higgs mechanism ({\it i.e.}, fundamental scalars developing a vev), the gluons 
acquire their (momentum-dependent) masses non-perturbatively, 
through the dynamical realization of the well-known Schwinger mechanism \cite{Schwinger:1962tn}. Despite the 
difference in their origin, the masses act in a very similar fashion at the 
level of the RG-invariant quantity 
associated with the corresponding gauge boson. 

Thus, in the case of the $W$-boson, the corresponding quantity would read (Euclidean momenta)
\be
{\widehat d}_{\chic W}(q^2) = \frac{ {\overline g}^2_{\chic W}(q^2)}{q^2+M^2_{\chic W}} 
\ee
with
\be
{\overline g}^2_{\chic W}(q^2) = 
g^2_{\chic W}(\mu)
\bigg[ 1+ b_{\chic W} g^2_{\chic W}(\mu) \int_0^1 dx \ln\left(\frac{q^2 x(1-x) + M^2_{\chic W}}{\mu^2}\right) - ...\bigg]^{-1} 
\ee
where $b_{\chic W}=11/24\pi^2$, and the ellipses denote the contributions of the fermion families.
Clearly, ${\widehat d}_{\chic W}(0) = {\overline g}^2_{\chic W}(0)/M^2_{\chic W}$, 
with 
${\overline g}^2_{\chic W}(0) = g^2_{\chic W}(\mu)
[1+ b_{\chic W}g^2_{\chic W}(\mu)  \ln (M^2_{\chic W}/\mu^2)]^{-1}$. 
Evidently, in the deep IR,  the coupling 
freezes at a constant value; 
Fermi's constant is in fact 
determined as $4 \sqrt{2} G_{\chic F}=  {\overline g}^2_{\chic W}(0)/M^2_{\chic W}$.
Note that in the case of QCD the corresponding combination,  ${\overline g}^2(0)/m^2(0)$
would be similar to a Nambu--Jona-Lasinio type of coupling~\cite{Nambu:1961tp}: at energies below the gluon mass $m$, the ``tree-level'' amplitude of four-quarks starts 
looking a lot like that of a four-Fermi interaction~\cite{f1}.  

This property of the ``freezing'' of the coupling can be reformulated 
in terms of what  in the language of the effective field theories is referred to as ``decoupling''~\cite{Appelquist:1974tg}. 
At energies sufficiently inferior to their masses, the particles appearing in the loops  
(in this case  the gauge bosons) 
seize to contribute to the ``running'' of the coupling.
Possibly large logarithmic constants, {\it e.g.}, $\ln (M^2_{\chic W}/\mu^2)$, may be reabsorbed in the 
renormalized value of the coupling. 
Of course, the ``decoupling'' as described above  should not be misinterpreted to mean 
that the running coupling vanishes; instead, as already mentioned, it freezes at a constant, non-zero value.
In other words: the ``decoupling'' does {\it not}  imply that the theory becomes free 
(non-interacting) in the IR.

\begin{center}
\begin{figure}[!t]
\includegraphics[scale=2.0]{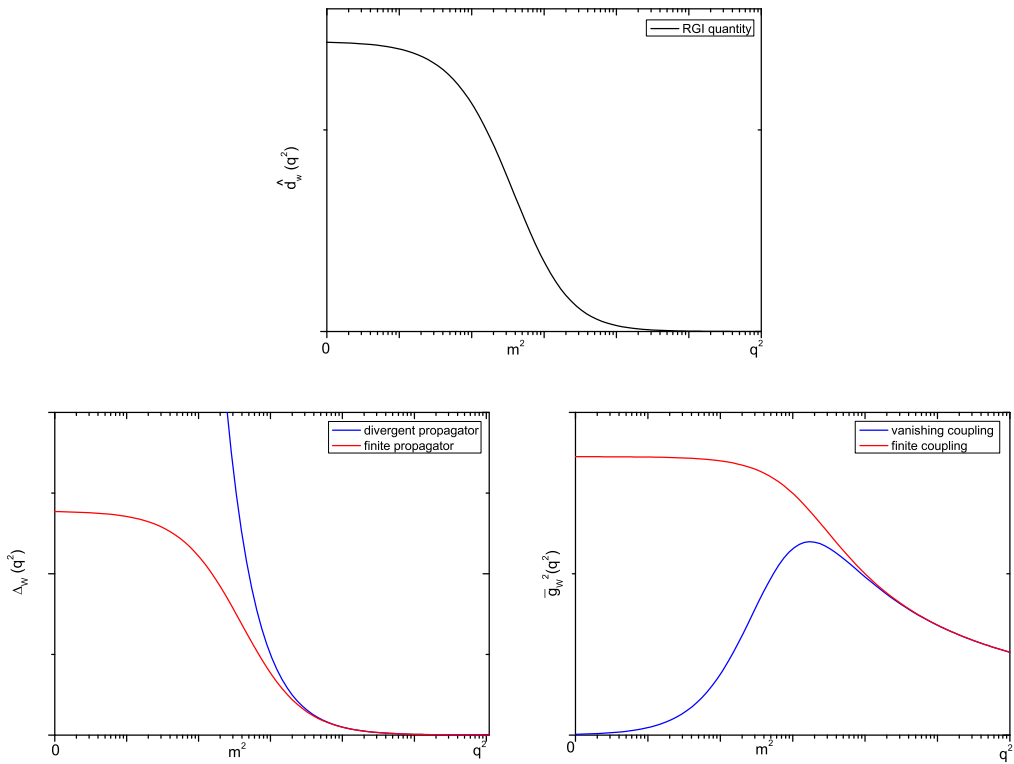}
\vspace{-1.5cm}
\caption{The same RG-invariant quantity decomposed in two different ways, one giving a 
divergent propagator and a vanishing coupling, and one giving a finite propagator and a finite coupling} 
\label{decomp}
\end{figure}
\end{center}

This last clarification is not without relevance for the question at hand, namely the 
definition of a physically meaningful effective charge. 
In particular, if one wants to 
extract an effective charge from an IR-finite gluon propagator 
(obtained from, {\it e.g.}, SD studies~\cite{Aguilar:2008xm} or from lattice simulations~\cite{Bernard:1981pg,Cucchieri:2007md,Bogolubsky:2007ud}),  
it would certainly be unwise to insist on the perturbative prescription, and 
simply factor out a $1/q^2$. Even though one is merely redistributing a given function, 
namely ${\widehat d}_{\chic W}(q^2)$,  
into two pieces, factoring out  $1/q^2$ deprives both of them of any physical meaning. 
Returning to the electroweak example, 
the effective coupling 
so defined would be given by 
the expression ${\widetilde g}^2_{\chic W}(q^2) = q^2 {\widehat d}_{\chic W}(q^2)$, 
and so ${\widetilde g}^2_{\chic W}(0) =0$; evidently,  
one would be attempting to describe weak interactions in terms of 
a massless, IR divergent gauge boson propagator and a vanishing effective coupling (See the curves in blue in Fig.~\ref{decomp}).     
Given that the gluon propagator is finite in the IR,  
if this latter (wrong) procedure were to be applied to QCD, it   
would furnish a completely unphysical coupling, 
namely one that vanishes in the deep IR, where QCD is expected to be (and is) strongly coupled.

As emphasized from the outset, the correct procedure is  
to factor out of the corresponding RG-invariant combination a ``massive'' propagator; 
in the PT case, we write the $\widehat{d}(q^2)$ of Eq.~(\ref{rgi}) 
\be
\widehat{d}(q^2) = \frac{\overline{g}^2(q^2)}{q^2 + m^2(q^2)}.
\label{ddef}
\ee
Given that \mbox{$\widehat{d}(q^2)=g^2 \widehat{\Delta}(q^2)$},
substituting Eq.~(\ref{ddef}) into (\ref{bqi2}) we obtain
\be
\alpha_{\chic{\mathrm{PT}}}(q^2)= [q^2+m^2(q^2)] \frac{\alpha(\mu^2) \Delta(q^2)}{[1+G(q^2)]^2},
\label{charge}
\ee
where we have used \mbox{$\alpha_{\chic{\mathrm{PT}}}(q^2)=\overline{g}^2(q^2)/4\pi$}.
As already mentioned, the dynamical mass  $m^2(q^2)$ appearing 
in the definition of $\alpha(q^2)$ is itself running;
the explicit form of this running will be discussed in Section~\ref{numan}.
Similarly, from the RG-invariant quantity defined starting from the ghost-gluon vertex, 
given in Eq.~(\ref{rg2}), we have that   
\be
\alpha_{\mathrm{gh}}(q^2) = \alpha'(\mu^2)(q^2+m^2(q^2))\Delta(q^2)F^2(q^2),
\label{rg3}
\ee
where $\alpha'(\mu^2)=\overline{g}_\mathrm{gh}(\mu^2)/4\pi$.

Since $\Delta(0)$,  $F(0)$, $G(0)$, and $m(0) \equiv m_0$ are all finite (non-vanishing), 
in the deep IR both couplings assume finite values given by 
\bea
\alpha_{\chic{\mathrm{PT}}}(0) &=& m^2_0 \alpha (\mu^2) \Delta(0) F^2(0)\,,
\nonumber\\
\alpha_{\mathrm{gh}}(0) &=& m^2_0 \alpha'(\mu^2) \Delta(0) [1+G(0)]^{-2}\,.
\label{coup0}
\eea

\section{\label{giSDE} Derivation of the identity from the dynamical equations}

In this section, we derive the central identity, valid  {\it only} in the Landau gauge, 
relating the ghost dressing function with 
a particular combination of the form-factors $G(q^2)$ and $L(q^2)$ 
appearing in the tensorial decomposition of $\Lambda_{\mu\nu}$ in Eq.~(\ref{LDec}). 
The proof hinges crucially on working in the 
Landau   gauge   ($\xi=0$),  where the  entire gluon   propagator $\Delta_{\mu\nu}(k)$ 
[and not just its self-energy $\Pi_{\mu\nu}(k)$] is transverse, {\it i.e.}, $k^{\mu} \Delta_{\mu\nu}(k) =0$. 
As we will see shortly, the operational consequence of this last property is that 
one can write $q^{\mu} \Delta_{\mu\nu}(k) =(q+k)^{\mu} \Delta_{\mu\nu}(k)$, thus generating  
for free the appropriate ghost-gluon vertex, as needed.

\subsection{Deriving the relation}

\begin{figure}[!t]
\includegraphics[width=11cm]{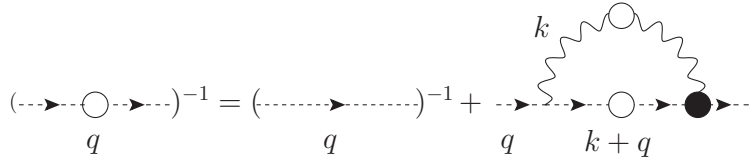}
\caption{The SDE for the ghost.}
\label{ghostSDE}
\end{figure}

The central relation is obtained as follows. 
First, consider the standard SD equation for the ghost propagator (Fig~\ref{ghostSDE}), 
\be
iD^{-1}(q^2) = q^2 +i g^2 C_{\rm {A}}  \int_k
\Gamma^{\mu}\Delta_{\mu\nu}(k)\gb^{\nu}(k,q) D(q+k).
\label{SDgh}
\ee
Then, contract both sides of
the defining equation~(\ref{LDec}) by the combination $q^{\mu}q^{\nu}$ to get
\be
[G(q^2) + L(q^2)]q^2 = g^2 C_{\rm {A}}
\int_k q_{\rho} \Delta^{\rho\sigma}(k)\, q^{\nu} H_{\sigma\nu}(k,q) D(k+q).
\label{s1}
\ee
Using Eq.~(\ref{qH}) and the transversality of the full gluon propagator, we can see that the rhs of Eq.~(\ref{s1}) 
is precisely the integral appearing in the ghost SDE~(\ref{SDgh}). Therefore 
\be
[G(q^2) + L(q^2)]q^2 =  iD^{-1}(q^2) - q^2,
\label{s2}
\ee
or, in terms of the ghost dressing function $F(q^2)$ [{\it viz}. Eq.~(\ref{ghostdress})]  
\be
1+ G(q^2) + L(q^2) = F^{-1}(q^2).
\label{funrel}
\ee
The relation of Eq.~(\ref{funrel}), derived here from the SDEs of the theory,  
has been first obtained in~\cite{Grassi:2004yq}, in the framework of the Batalin-Vilkovisky quantization formalism.  
As was shown there, the relation is a direct consequence of the fundamental BRST symmetry.

Let us study the functions $G(q^2)$ and  $L(q^2)$ more closely. From Eq.~(\ref{LDec}) we have that (in $d$ dimensions) 
\be
G(q^2) = \frac{1}{(d-1)q^2} \left(q^2 \Lambda_{\mu}^{\mu} - q^{\mu}q^{\nu}\Lambda_{\mu\nu} 
\right),\qquad L(q^2) = \frac{1}{(d-1)q^2} \left(d q^{\mu}q^{\nu}\Lambda_{\mu\nu} - q^2\Lambda_{\mu}^{\mu} \right),
\label{s3}
\ee 
which then gives, in terms of the SDE integrals
\bea
G(q^2) &=& \frac{g^2 C_{\rm {A}}}{d-1}
\left[ 
\int_k \Delta^{\rho\sigma}(k)\, H_{\sigma\rho}(k,q) D(k+q)
+i
\frac{1}{q^2} \int_k q^{\rho} \Delta_{\rho\sigma}(k)\, {\gb}^{\sigma}(k,q) D(k+q)
\right],
\nonumber\\
L(q^2) &=& -\frac{g^2 C_{\rm {A}}}{d-1}
\left[i
\frac{d}{q^2} \int_k q^{\rho} \Delta_{\rho\sigma}(k)\, {\gb}^{\sigma}(k,q) D(k+q)
\!+\!\! \int_k \Delta^{\rho\sigma}(k)\, H_{\sigma\rho}(k,q) D(k+q)
\right]\!.
\label{s4}
\eea 

Inserting the decomposition of Eq.~(\ref{Gtens}) and Eq.~(\ref{Htens})
into  Eq.~(\ref{s4}), and setting  
\be
f(k,q) \equiv  \frac{(k \cdot q)^2}{k^2 q^2} \,,
\ee
we obtain 
\bea
G(q^2)\! &=&\! \frac{g^2 C_{\rm {A}}}{d-1}\!
\int_k\! \left\{(d-1) A_1(k,q) -  [1-f(k,q)]\left[B_1(k,q)-q^2 A_2(k,q)\right]\right\} \Delta (k)  D(k+q),
\nonumber\\
L(q^2)\! &=&\!  \frac{g^2 C_{\rm {A}}}{d-1}\!
\int_k\! \left\{(1-d) A_1(k,q) + [1-f(k,q)] \left[dB_1(k,q)
-q^2A_2(k,q)\right]\right\}
\Delta (k)  D(k+q),
\nonumber\\
&&{}
\label{tt1}
\eea 
while from Eq.~(\ref{SDgh}) 
\be
F^{-1}(q^2) = 1 +g^2 C_{\rm {A}} \int_k\, [1-f(k,q)] B_1(k,q)\Delta (k)  D(k+q).
\label{tt2}
\ee
Clearly, Eq.~(\ref{funrel}) is automatically satisfied. 

\subsection{Renormalization}

Of course, all quantities appearing in Eq.~(\ref{tt1}) and  Eq.~(\ref{tt2}) are unrenormalized 
(we have suppressed the corresponding subscript ``0'' for simplicity); 
in particular, Eq.~(\ref{funrel}) involves unrenormalized $G(q^2)$, $L(q^2)$, and $F(q^2)$.  
It is easy to recognize, for example,  by substituting in the corresponding 
integrals tree-level expressions,
that  $F^{-1}(q^2)$ and  $G(q^2)$ have the same 
leading dependence on the UV cutoff $\Lambda_\chic{\mathrm{UV}}$, namely 
\be
F^{-1}_{\mathrm{UV}} (q^2) = G_{\mathrm{UV}} (q^2) = 
\frac{3g^2 C_{\rm {A}}}{64 \pi^2} \ln\left(\frac{\Lambda^2_\chic{\mathrm{UV}}}{q^2}\right), 
\label{uvdiv}
\ee
while $L(q^2)$ is finite (independent of $\Lambda_\chic{\mathrm{UV}}$) at leading order.
The next step is therefore to carry out the necessary renormalization.

As already mentioned above, the origin of the basic relation of Eq.~(\ref{funrel}) 
is the BRST symmetry of the theory; in that sense, Eq.~(\ref{funrel}) has the same origin 
as the Slavnov-Taylor identities of the theory. Therefore, just as happens with 
the Slavnov-Taylor identities, Eq.~(\ref{funrel}) should not be deformed after renormalization.
Of course, the prototype example of such a situation are the Ward identities of QED; the requirement that 
the fundamental Ward identity $q^{\mu}\Gamma_{\mu} = S^{-1}(p+q)-S^{-1}(p)$ should retain the same 
form before and after renormalization leads to the well-known textbook relation $Z_1=Z_2$ 
between the corresponding renormalization constants \cite{Itzykson:rh}.  
Similarly, for the case at hand, the renormalization must be carried out in such a way as to 
preserve the form Eq.~(\ref{funrel}).  
Specifically, using the definition given in Eq.~(\ref{Lamrel}), 
in order to preserve the relation~(\ref{funrel}) after renormalization, we must impose that 
\be
Z_\Lambda = Z_{c}.
\label{renconst3}
\ee
In addition, by virtue of (\ref{qH}), and for the same reason explained above,  
we have that, in the Landau gauge ${\gb}_{\nu}(k,q)$ and $H_{\sigma\nu}(k,q)$ 
must be renormalized by the same renormalization constant, namely $Z_1$ 
[{\it viz.} Eq.~(\ref{renconst})]; for the Taylor kinematics, we have that $Z_1=1$ 
[see Eq.~(\ref{Z1T})].

Then, it is straightforward to renormalize Eq.~(\ref{tt1}) and Eq.~(\ref{tt2}); using 
\bea 
F^{-1}(q^2,\mu^2)&=&Z_c(\Lambda^2_\chic{\mathrm{UV}}, \mu^2) F_0^{-1} (q^2,\Lambda^2_\chic{\mathrm{UV}}), \nonumber\\ 
1+G(q^2,\mu^2)&=&Z_c(\Lambda^2_\chic{\mathrm{UV}}, \mu^2)[1+G_0(q^2,\Lambda^2_\chic{\mathrm{UV}})], \nonumber\\ 
L(q^2,\mu^2)&=& Z_c(\Lambda^2_\chic{\mathrm{UV}}, \mu^2) L_0(q^2,\Lambda^2_\chic{\mathrm{UV}}),
\label{Zcren}
\eea
we have that 
\be
F^{-1}(q^2) = Z_c +g^2 C_{\rm {A}} \int_k\, [1-f(k,q)] B_1(k,q)\Delta (k)  D(k+q) \,,
\label{Fren}
\ee
and 
\be
1+G(q^2) = Z_c + \frac{g^2 C_{\rm {A}}}{d-1}\!
\int_k\!\! \left\{\!(d-1) A_1(k,q) -  [1-f(k,q)]\left[B_1(k,q)-q^2 A_2(k,q)\right]\!\right\}\! \Delta (k)  D(k+q),
\label{Gren}
\ee
while the equation for  $L(q^2)$ remains unchanged, {\it i.e.}, one simply 
replaces in the second equation of  (\ref{tt1}) the unrenormalized quantities by renormalized ones.
This is consistent with the general observation made in~\cite{Grassi:2004yq}, according to which 
$L(q^2)$ does need its own counterterm, {\it i.e.}, one proportional to $q_{\mu} q_{\nu}$, in order to get renormalized. 
The situation is similar to what happens with the $\sigma_{\mu\nu}q^{\nu}$ part of the standard QED vertex:
The renormalizability of the theory forbids of course a counterterm proportional to such a tensorial structure; 
the magnetic form factor (usually denoted by $F_2(q^2)$) is made finite (beyond one loop) after multiplication by 
the renormalization constant $Z_1$ (whose counterterms are  proportional to $\gamma_{\mu}$). 
Thus, while the one-loop answer for $F_2$ is finite, 
at higher orders one gets divergences proportional to $\sigma_{\mu\nu}q^{\nu}$
which are, however, canceled exactly (order by order) by the inclusion of the $Z_1$ 
counterterms in the Feynman graphs of the previous order.    
For this reason, just as $F_2$, despite its one-loop finiteness 
$L$ depends in general on the UV cutoff $\Lambda^2_\chic{\mathrm{UV}}$, as indicated explicitly in Eq.~(\ref{Zcren}).

\subsection{Calculations and approximations}

In order to study the relevant equations further,
we will approximate the form factors $A_1(k,q)$ and $B_1(k,q)$ with their 
tree-level values, {\it i.e.}, $A_1(k,q)=  B_1(k,q)=1$, and \mbox{$A_2(k,q)=0$}; 
according to lattice 
studies~\cite{Cucchieri:2004sq}, this appears to be a very good approximation.  
Then, we obtain from Eqs.~(\ref{tt1}) and~(\ref{tt2})
\bea
F^{-1}(q^2) &=& Z_c +g^2 C_{\rm {A}} \int_k [1-f(k,q)] \Delta (k)  D(k+q),
\nonumber\\
1+G(q^2) &=& Z_c + \frac{g^2 C_{\rm {A}}}{d-1}\int_k [
(d-2)+ f(k,q)]\Delta (k)  D(k+q),
\nonumber\\
L(q^2) &=& \frac{g^2 C_{\rm {A}}}{d-1}\int_k 
[1 - d \,f(k,q)]\Delta (k)  D(k+q).
\label{simple}
\eea 
Now, it turns out that if $F$ and $\Delta$ are both IR finite,   
then
\be
\left.\int_k [1-d \,f(k,q)]\Delta(k)D(k+q) \right|_{q\to 0} = 0,
\label{id1}
\ee
To see this, one may use the result 
$\int_k k_{\mu}k_{\nu} F(k) \Delta(k) = g_{\mu\nu} d^{-1} \int_k F(k) \Delta(k)$, 
or, equivalently, go to spherical coordinates and use that\footnote{
Recall that $\int_{0}^{\pi} \!\!\! d\theta\sin^n\theta= 
\frac{\Gamma\left(\frac{n+1}{2}\right) \Gamma\left( \frac{1}{2}\right) }{\Gamma\left( \frac{n+2}{2}\right) }$.}
\be 
\int_{0}^{\pi} \!\!\! d\theta\sin^d\theta(1-d \cos^2\theta)=0.
\label{angint}
\ee
Thus, from Eq.~(\ref{simple}) we obtain the important result 
\be
L(0)= 0, 
\label{L0IR}
\ee
under the assumption that  $F$ and $\Delta$ are IR finite.
In addition, using (\ref{id1}), we obtain 
\be
F^{-1}(0)  = 1+ G(0) =  Z_c + \frac{g^2 C_{\rm {A}}(d-1)}{d} \int_k \Delta (k) D(k).
\label{t3}
\ee

Note that perturbatively, at one loop, 
Eq.~(\ref{id1}) does {\it not} hold, because in that case $\Delta(k)$ is {\it not} IR finite; 
consequently, at one loop $L(0)\neq 0$. 
Specifically in this case, using dimensional regularization, 
we obtain the $q$-independent result
\be
\int_k \frac{1- d\, f(k,q)}{k^2(k+q)^2} = - \frac{3}{2} \frac i{16\pi^2},
\label{id2}
\ee
which gives 
\be
L_0(q^2) = \frac{g^2 C_{\rm {A}}}{32\pi^2}.
\label{L0}
\ee
If instead we were to use an IR finite gluon propagator, modeled simply by \mbox{$\Delta^{-1}(k) = k^2-m^2$}, 
the same calculation would show that $L_m (q^2)$ depends non-trivially on $q^2$ 
[see Eq.~(\ref{Lmx}) below], and in fact, $L_m (0) = 0$.

We next go to the Euclidean space, by 
setting $-q^2=q^2_\mathrm{E}$, and defining $\Delta_\mathrm{E}(q^2_\mathrm{E})=-\Delta(-q^2_\mathrm{E})$, $D_\mathrm{E}(q^2_\mathrm{E})=-D(-q^2_\mathrm{E})$, and for the integration measure $\int_k=i\int_{k_\mathrm{E}}$. Then,  using Eq.~(\ref{ghostdress}) and suppressing the subscript ``E'', we obtain from Eqs.~(\ref{tt1}) and~(\ref{tt2})
\bea
F^{-1}(q^2) &=& Z_c -g^2 C_{\rm {A}} \int_k [1- f(k,q)] \Delta (k)  D(k+q),
\nonumber\\
1+G(q^2) &=& Z_c -\frac{g^2 C_{\rm {A}}}{d-1}\int_k [(d-2)+ f(k,q)] \Delta (k)  D(k+q),
\nonumber\\
L(q^2) &=& -\frac{g^2 C_{\rm {A}}}{d-1}\int_k [1 - d\,f(k,q)] \Delta (k)  D(k+q).
\label{simple1}
\eea 
Next let us introduce spherical coordinates. Setting $q^2=x$, $k^2=y$, 
we have that \mbox{$k \cdot q =\sqrt{xy}\cos\theta$}, and so $(k \cdot q)^2/q^2 =y\cos^2\theta $, and 
$(k+q)^2 = x + y +2 \sqrt{xy}\cos\theta$. Moreover, at $d=4$,
the measure is given by 
\be
\int d^{4} k = 
2 \pi\!\!\int_{0}^{\pi} \!\!\! d\theta\sin^2\theta\, 
\int_{0}^{\infty}\!\!\! dy\ y.
\label{spher}
\ee

Let us first consider the case in which the ghost propagator assumes its tree-level form, namely 
$D(k+q)=1/(k+q)^2$. Then, using the results
\bea
\int_0^\pi\!d\theta\frac{\sin^2\theta}{x + y +2 \sqrt{xy}\cos\theta}&=& 
\frac{\pi}2\left[\frac1x\Theta(x-y)+\frac1y\Theta(y-x)\right],\nonumber \\
\int_0^\pi\!d\theta\frac{\sin^2\theta\cos^2\theta}{x + y +2 \sqrt{xy}\cos\theta} 
&=&\frac{\pi}8\left[\frac1x\bigg(1+\frac yx\bigg)\Theta(x-y)+\frac1y\bigg(1+\frac xy\bigg)\Theta(y-x)\right],
\eea
where $\Theta(x)$ is the Heaviside step function, one obtains
\bea
1+G(x) &=&  Z_c - \frac{\alpha_s C_{\rm {A}}}{16\pi} \left[
\frac{1}{x}\int_{0}^{x}\!\!\! dy\  y \left(3 + \frac{y}{3x}\right) \Delta(y) 
+ \int_{x}^{\infty}\!\!\! dy \left(3 + \frac{x}{3y}\right)\Delta(y) 
\right],
\nonumber\\
L(x) &=&  \frac{\alpha_s C_{\rm {A}}}{12\pi} \left[
\frac{1}{x^2}\int_{0}^{x}\!\!\! dy\ y^2 \Delta(y) 
+ x \int_{x}^{\infty}\!\!\! dy \frac{\Delta(y)}{y}
\right],
\nonumber\\
F^{-1}(x) &=&  Z_c- \frac{\alpha_s C_{\rm {A}}}{16\pi} \left[
\frac{1}{x}\int_{0}^{x}\!\!\! dy\  y \left(3 - \frac{y}{x}\right) \Delta(y) 
+ \int_{x}^{\infty}\!\!\! dy \left(3 - \frac{x}{y}\right)\Delta(y) 
\right].
\label{spherres1}
\eea
Substituting into the equation for 
 $L(x)$ the tree-level value for $\Delta(y)$ we obtain  
the constant result $L_0(x)$ of Eq.~(\ref{L0}). On the other hand, using $\Delta(y) = (y+m^2)^{-1}$, we find 
\be
L_m(x) = \frac{\alpha_s C_{\rm {A}}}{12\pi}  \left\{
\frac{1}{x^2}\left[\frac{x^2}{2} - m^2 x +  m^4 \ln\left(1+\frac{x}{m^2}\right)\right]  
+ \frac{x}{m^2} \ln\left(1+\frac{m^2}{x}\right) \right\},
\label{Lmx}
\ee
from which we clearly see that $L_m(0) = 0$. In addition, for large $x$, 
$L_m(x)$ goes over to the massless limit of Eq.~(\ref{L0}). 

The general case for an arbitrary ghost dressing function $F(k+q)$, can be treated by means of the angular approximation. 
Specifically, one write approximately 
\bea
1+G(x) &=&  Z_c - \frac{\alpha_s C_{\rm {A}}}{16\pi}\left[
\frac{F(x)}{x}\int_{0}^{x}\!\!\! dy\  y \left(3 + \frac{y}{3x}\right) \Delta(y) 
+ \int_{x}^{\infty}\!\!\! dy \left(3 + \frac{x}{3y}\right)\Delta(y)F(y) 
\right],
\nonumber\\
L(x) &=&  \frac{\alpha_s C_{\rm {A}}}{12\pi} \left[
\frac{F(x)}{x^2}\int_{0}^{x}\!\!\! dy\ y^2 \Delta(y) 
+ x \int_{x}^{\infty}\!\!\! dy \frac{\Delta(y) F(y)}{y}
\right],
\nonumber\\
F^{-1}(x) &=& Z_c - \frac{\alpha_s C_{\rm {A}}}{16\pi} \left[
\frac{F(x)}{x}\int_{0}^{x}\!\!\! dy\  y \left(3 - \frac{y}{x}\right) \Delta(y) 
+ \int_{x}^{\infty}\!\!\! dy \left(3 - \frac{x}{y}\right)\Delta(y) F(y) 
\right].
\label{LFGr}
\eea
It is then easy to see ({\it e.g.}, by means of the change of variables $y=zx$) that if $\Delta$ and $F$ are 
IR finite, then $L(0)=0$, as claimed before.
Let us now assume that the renormalization condition for $F(x)$ was chosen 
to be \mbox{$F(\mu^2) =1$}. This condition, when inserted into the third equation of (\ref{LFGr}), 
allows one to express $Z_c$ as  
\be
Z_c = 1+ \frac{\alpha_s C_{\rm {A}}}{16\pi}  \left[
\frac{1}{\mu^2}\int_{0}^{\mu^2}\!\!\! dy  y \left(3 - \frac{y}{\mu^2}\right) \Delta(y) 
+ \int_{\mu^2}^{\infty}\!\!\! dy \left(3 - \frac{\mu^2}{y}\right)\Delta(y) F(y) 
\right],
\label{Zcexp}
\ee
and may be used to cast (\ref{LFGr}) into a manifestly renormalized form.
Note that if one choses $F(\mu^2) =1$ then one cannot choose simultaneously  
$G(\mu^2)=0$, because that would violate the identity of Eq.~(\ref{funrel}), 
given that $L(\mu^2)\neq 0$. In fact, once $F(\mu^2) =1$ has been imposed, 
the value of $G(\mu^2)$ is completely determined from its own equation, {\it i.e.}, 
the first equation in~(\ref{LFGr}). 

In addition in the MOM scheme the conventional and PT propagator cannot be made equal at the renormalization point, since the identity~(\ref{bqi2})  implies\linebreak
$\widehat{\Delta}^{(-1)}(\mu^2) = \mu^{2}\left[ 1+G^2(\mu^2) \right]^{2}$.

\subsection{Implications for the effective charges}

After this general discussion, let us now return to the couplings, and 
discuss the implications of the identity and the dynamics 
we have derived.

First of all, comparing Eq.~(\ref{ptrgi}) and Eq.~(\ref{rg2}), it is clear that $g(\mu)=g^{\prime}(\mu)$,
by virtue of Eq.~(\ref{renconst3}). Therefore, using Eq.~(\ref{bqi2}), one can get a relation between the two RG-invariant quantities, 
$\widehat{r}(q^2)$ and $\widehat d(q^2)$, namely
\be
\widehat{r}(q^2)=[1+G(q^2)]^2 F^2(q^2)\widehat d(q^2).
\label{rel_rgi}
\ee
From this last equality follows that $\alpha_{\chic{\mathrm{PT}}}$ and $\alpha_{\mathrm{gh}}(q^2)$ 
are related by 
\be
\alpha_{\mathrm{gh}}(q^2) = [1+G(q^2)]^2 F^2(q^2)\alpha_{\chic{\mathrm{PT}}}(q^2),
\label{coup_charge}
\ee
After using Eq.~(\ref{funrel}), we have that  
\be
\alpha_{\mathrm{gh}}(q^2)=\left[\frac{1+G(q^2)}{1+G(q^2)+L(q^2)}\right]^2 
\alpha_{\chic{\mathrm{PT}}}(q^2).
\label{relcoup}
\ee
or, equivalently,
\be
\alpha_{\chic{\mathrm{PT}}}(q^2) = \alpha_{\mathrm{gh}}(q^2)\left[1+ \frac{L(q^2)}{1+G(q^2)}\right]^2 \,.
\label{relcoup2}
\ee
Evidently, the two couplings can only coincide at two points:
(i) at $q^2=0$, where, 
due to the fact that $L(0)=0$ [see Eq.~(\ref{L0IR})], 
we have that 
\be
\alpha_{\mathrm{gh}}(0) = \alpha_{\chic{\mathrm{PT}}}(0), 
\ee
and (ii) at $q^2= \infty$,
given that in the deep UV $L(q^2)$ approaches a constant.
Note in fact that  
the two effective charges {\it cannot} coincide 
at the renormalization point $\mu$, where  
\be
\alpha_{\mathrm{gh}}(\mu^2) = [1-L(\mu^2)]^2 \alpha_{\chic{\mathrm{PT}}}(\mu^2);
\ee
this can be understood also in terms of the discussion following Eq.~(\ref{Zcexp}). 

As we will see in the next section, the numerical analysis reveals that 
$L(q^2)$ is fairly small compared to $G(q^2)$; thus, 
even in the region of intermediate momenta, where the 
difference reaches its maximum, 
the relative difference between the two charges is less than 5$\%$.

\section{\label{numan}Numerical analysis} 

In this section we will compute the QCD effective charges defined above, 
using as input for the various Green's functions appearing in their definitions 
the non-perturbative solutions of the corresponding SDEs, in the Landau gauge.  
In particular, we will solve numerically a system of three coupled non-linear integral equations, 
containing $\Delta(q^2)$, $F(q^2)$, and $G(q^2)$ as unknown quantities. 
Once solutions for these three functions have been obtained, then $L(q^2)$
is fully determined by its corresponding equation, namely the second one in Eq.~(\ref{LFGr}). 

\subsection{The system of SD equations}

\begin{figure}[!t]
\includegraphics[width=15cm]{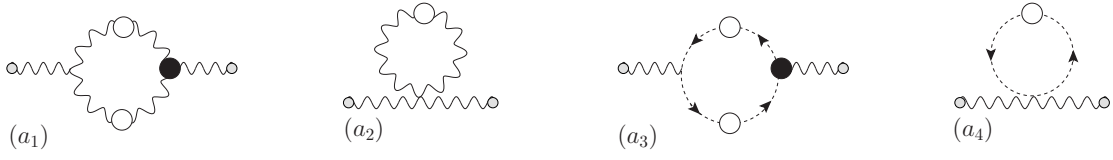}
\caption{The new  SDE for the gluon propagator at the one-loop dressed level.}
\label{SDeqs}
\end{figure}

The two SDEs determining  $F(q^2)$ and  $G(q^2)$ are given in Eq.~(\ref{LFGr}).
The SD equation governing $\Delta(q^2)$, is given by~\cite{Aguilar:2008xm} 
\be
[1+G(q^2)]^2\Delta^{-1}(q^2)P_{\mu\nu}(q) = 
q^2 P_{\mu\nu}(q) + i\sum_{i=1}^{4}(a_i)_{\mu\nu},
\label{SDgl}
\ee
where the diagrams $(a_i)_{\mu\nu}$ are shown in Fig.~\ref{SDeqs}. As explained in~\cite{Aguilar:2008xm}, 
due to the abelian Ward-identities satisfied by the fully-dressed vertices in the PT-BFM scheme, 
we have that $q^{\mu}[(a_1)_{\mu\nu}+ (a_2)_{\mu\nu}] = q^{\mu}[(a_3)_{\mu\nu}+ (a_4)_{\mu\nu}] =0$.
This last property enforces the transversality of the gluon self-energy ``order-by-order'' in the dressed-loop expansion, which is one of the central features of the gauge-invariant 
Schwinger-Dyson truncation scheme defined within the PT-BFM framework~\cite{Binosi:2007pi}. 

After introducing appropriate Ans\"atze for the aforementioned  fully-dressed vertices, 
we finally arrive at the integral equation
\bea
[1+G(q^2)]^2 \Delta^{-1}(q^2)\, &=&\, 
q^2 - \frac{g^2C_A}{6} \left[ 
\int_k \!\!\Delta(k)\Delta(k+q)f_1 
+ \int_k  \!\!\Delta(k) f_2 
- \frac{1}{2} \int_k \frac{q^2}{k^2 (k+q)^2}\right] 
\nonumber\\
&+&  g^2 C_A \bigg[\frac{4}{3}
\int_k \left[ k^2 - \frac{(k\cdot q)^2}{q^2}\right] D(k) D(k+q)
- 2 \int_k  D(k)\bigg],
\label{sdef}
\eea
with
\bea
f_1 &=& 20q^2 + 18k^2 -6(k+q)^2 + \frac{(q^2)^2}{(k+q)^2}-  (k\cdot q)^2\bigg[ \frac{20}{k^2}
+ \frac{10}{q^2} + \frac{q^2}{k^2 (k+q)^2}
+\frac{2 (k+q)^2}{q^2 k^2}\bigg] \,,\nonumber\\
f_2 &=& -\frac{27}{2} -8 \frac{ k^2}{(k+q)^2}
+8 \frac{q^2}{(k+q)^2} 
+ 4 \frac{(k\cdot q)^2}{k^2(k+q)^2}
- 4 \frac{(k\cdot q)^2}{q^2(k+q)^2},\
\label{f1f2}
\eea
The important point is that, by virtue of the massless composite poles introduced into the SDE through the 
particular Ans\"atze employed~\cite{Jackiw:1973tr, Cornwall:1982zr,Aguilar:2006gr}, 
one obtains an IR finite solution for the gluon propagator, {\it i.e.}, a solution with  $\Delta^{-1}(0) > 0$, 
in complete agreement with a large body of lattice data~\cite{Bernard:1981pg, Cucchieri:2007md, Bogolubsky:2007ud}. 
As explained in detail in~\cite{Aguilar:2008xm}, the formal expression determining 
$\Delta^{-1}(0)$ involves quadratically divergent integrals, which may be  
regulated using the standard rules of dimensional regularization. This procedure  
leaves the (finite) value of  $\Delta^{-1}(0)$ largely undetermined; therefore, 
in practice, $\Delta^{-1}(0)$ is treated as a free parameter, whose value is to be fixed 
using phenomenological constraints or lattice data. 
In addition, and because $\Delta^{-1}(0)$ is finite, the ghost dressing function $F(q^2)$ clearly saturates 
in the deep IR, reaching a finite value at $q^2=0$ (no ``enhancement'' observed), 
in agreement with recent lattice data \cite{Cucchieri:2007md,Bogolubsky:2007ud}, and a variety of independent studies~\cite{Boucaud:2006if,Dudal:2008sp}.

\begin{figure}[!t]
\begin{minipage}[b]{0.5\linewidth}
\centering
\hspace{-1cm}
\includegraphics[scale=0.8]{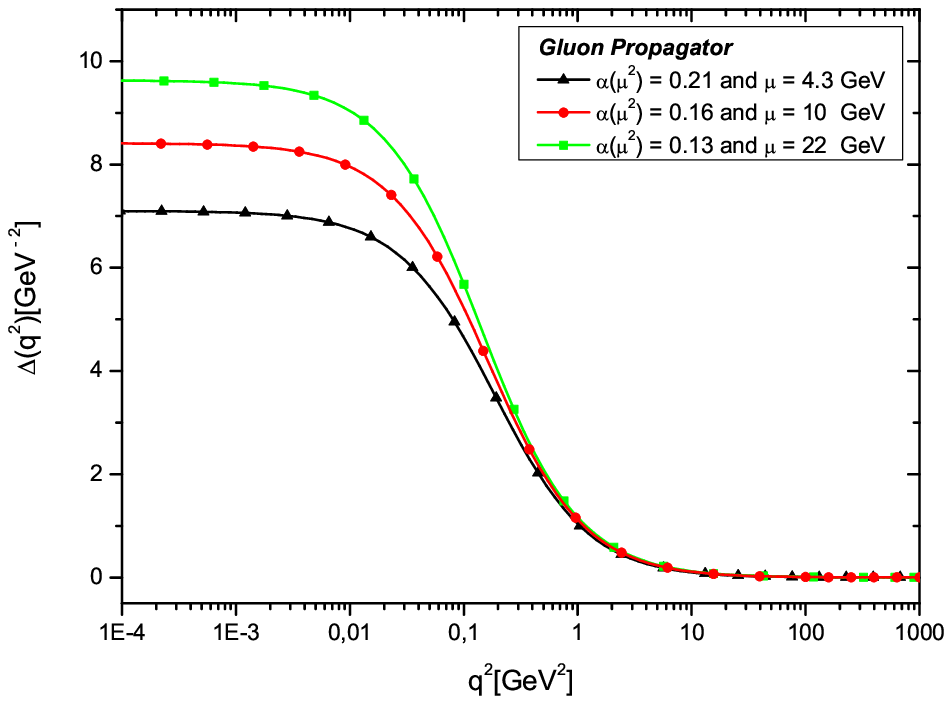}
\end{minipage}
\hspace{0.5cm}
\begin{minipage}[b]{0.45\linewidth}
\centering
\hspace{-2.0cm}
\includegraphics[scale=0.8]{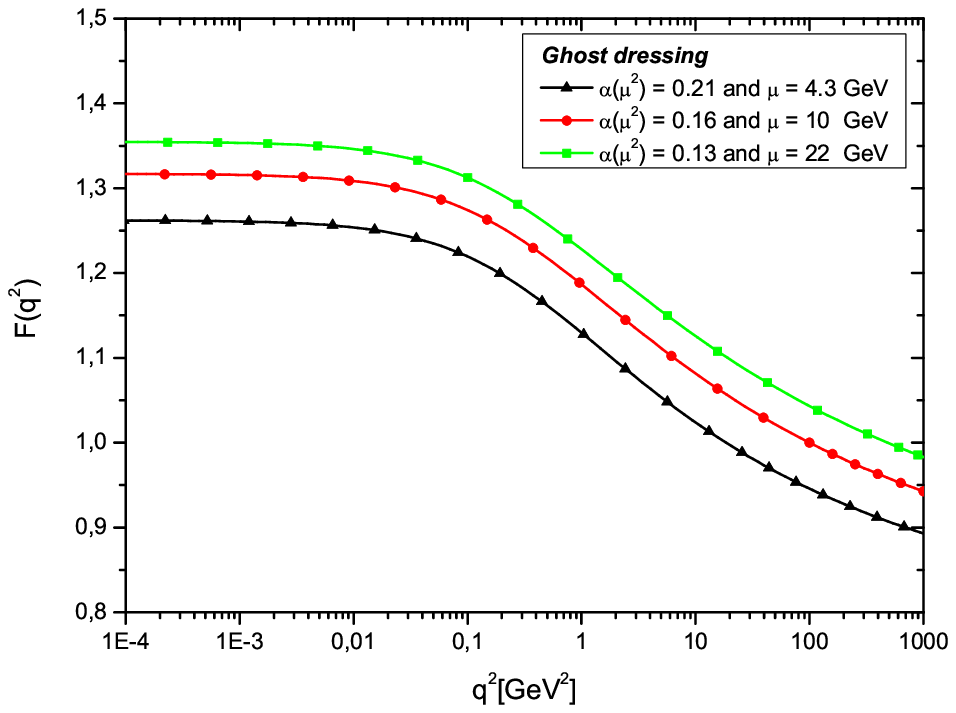}
\end{minipage}
\vspace{-1.5cm}
\caption{{\it Left panel}: Numerical solutions for the gluon propagator obtained from the SDE
using three different renormalization points: $\mu = 4.3 \,\mbox{GeV}$ and \mbox{$\alpha(\mu^2)=0.21$} (black curve), $\mu = 10 \,\mbox{GeV}$ and \mbox{$\alpha(\mu^2)=0.16$} (red curve), $\mu = 22 \,\mbox{GeV}$ and \mbox{$\alpha(\mu^2)=0.13$} (green curve). {\it Right panel}: 
The ghost dressing function $F(q^2)$ obtained from its corresponding SDE and renormalized at the same points.}
\label{fig2}
\end{figure}

\subsection{Solutions and checks}
 
In Fig.~\ref{fig2}, we show the numerical results for $\Delta(q^2)$ and $F(q^2)$,  renormalized at three different points.  On the left panel, the black curve represents the numerical solution of $\Delta(q^2)$ when  \mbox{$\alpha(\mu^2)=0.21$} and $\mu = 4.3 \,\mbox{GeV}$. 
The red curve is obtained when  \mbox{$\alpha(\mu^2)=0.16$} and $\mu = 10 \,\mbox{GeV}$, 
while for the green curve  we used \mbox{$\alpha(\mu^2)=0.13$} and $\mu = 22 \,\mbox{GeV}$.
On the right panel we plot the corresponding $F(q^2)$ renormalized at the same points.     

\begin{figure}[!t]
\begin{minipage}[b]{0.5\linewidth}
\centering
\hspace{-1cm}
\includegraphics[scale=0.8]{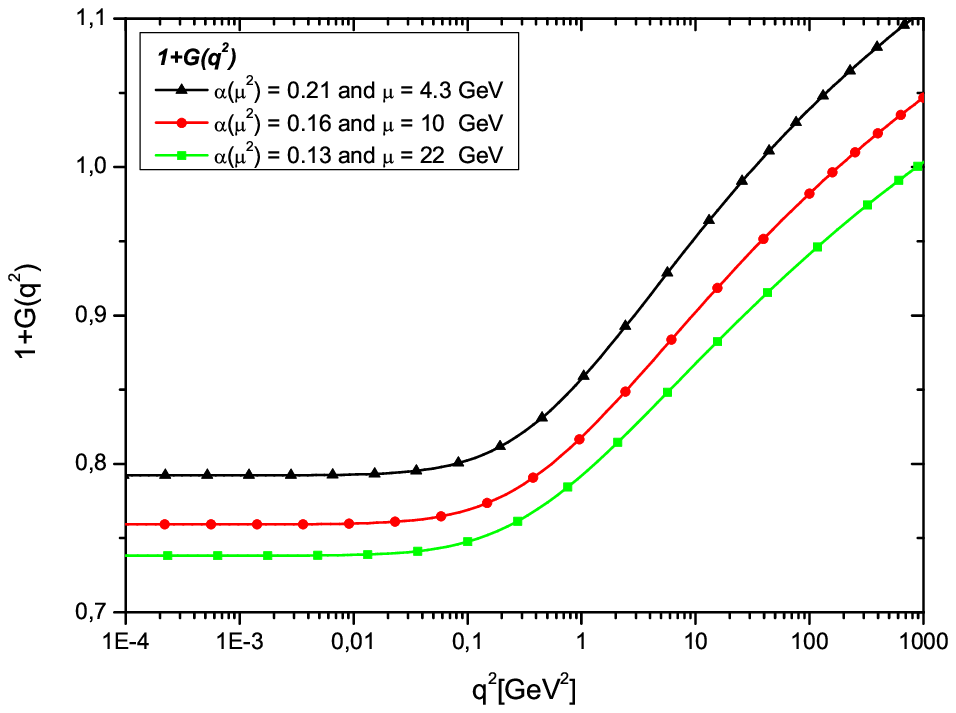}
\end{minipage}
\hspace{0.5cm}
\begin{minipage}[b]{0.45\linewidth}
\centering
\hspace{-2.0cm}
\includegraphics[scale=0.8]{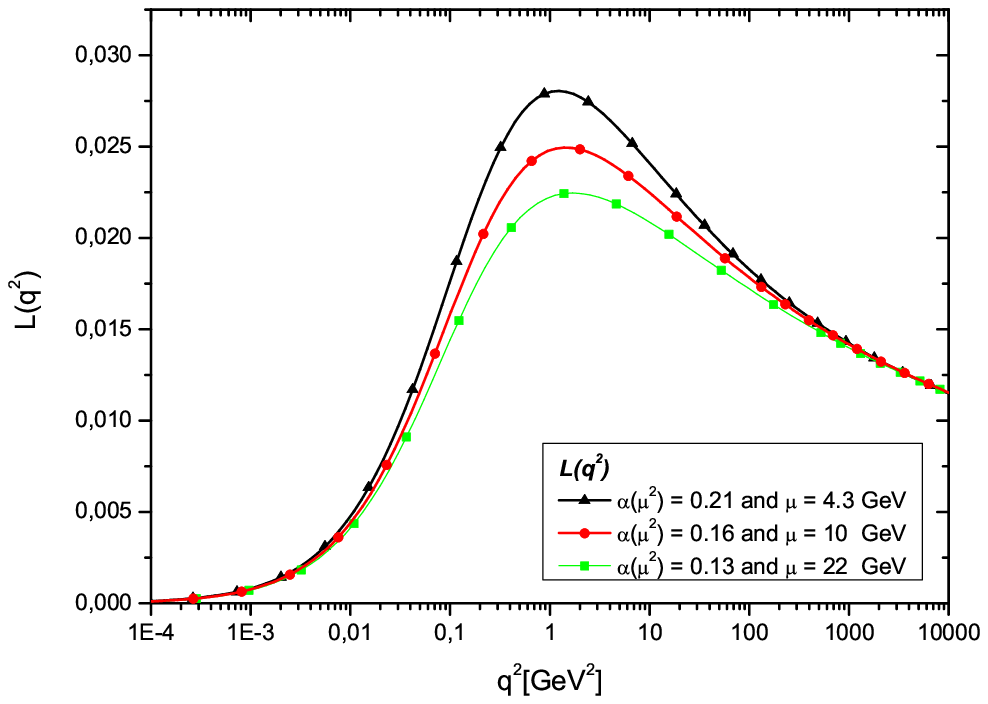}
\end{minipage}
\vspace{-1.5cm}
\caption{{\it Left panel}: $1+G(q^2)$ determined from Eq.~(\ref{LFGr}), using the solutions
for $\Delta(q^2)$ and $D(q^2)$ presented in the Fig.~\ref{fig2} at the same renormalization point. {\it Right panel}: 
The function $L(q^2)$ obtained from Eq.~(\ref{LFGr}).}
\label{fig3}
\end{figure}

In Fig.~\ref{fig3} we show the numerical results for the 
functions $1+G(q^2)$ and $L(q^2)$,  using the same renormalization points used previously. The color pattern is also the same as before. 
For values of $q^2<0.1 \mbox{GeV}^2$, we then see that $[1+G(q^2)]^2$ develops a plateau and saturates at a finite value in the deep IR region. In the UV region, we instead recover the perturbative behavior~(\ref{pert_gluon}). 
On the other hand, $L(q^2)$ (right panel) shows a maximum in the intermediate momentum region, while, as expected, $L(0)=0$.

\begin{figure}[!b]
\begin{center}
\includegraphics[scale=0.8]{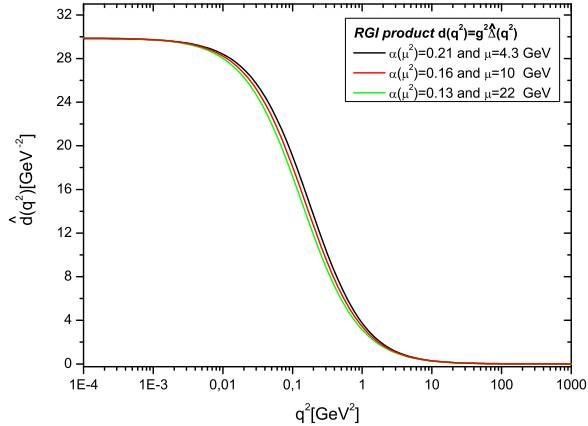}
\end{center}
\caption{The product $\widehat{d}(q^2)$ obtained combining the results for $\Delta(q^2)$ and $[1+G(q^2)]^2$
according to Eq.~(\ref{rgi}).}
\label{fig4}
\end{figure}

With all ingredients defined, the first thing one can check is whether Eq.~(\ref{rgi}) 
gives rise to a RG-invariant combination, as expected. Using the latter definition,  we can combine  
the different data sets  for $\Delta(q^2)$ and $[1+G(q^2)]^2$ at different renormalization points, 
to arrive at  the curves shown in Fig.~\ref{fig4}.  Indeed, we see that the combination $\widehat{d}(q^2)$ 
is practically independent of the renormalization point chosen.

\begin{figure}[!t]
\begin{center}
\includegraphics[scale=0.9]{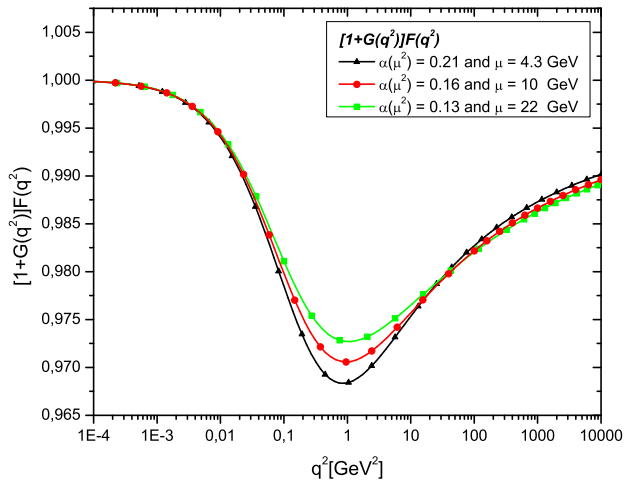}
\end{center}
\caption{The product $(1+G) F$ for different values of the renormalization point $\mu$. Note the fine scale of the $y$-axis.}
\label{fig10}
\end{figure}

In addition, from the available solutions we can compute the product \mbox{$(1+G) F$}, 
which, according to Eq.~(\ref{coup_charge}), relates the two effective charges of interest.  
Evidently, since both effective charges are supposed to be RG-invariant quantities, so 
should be the product \mbox{$(1+G) F$} relating them. In Fig.~\ref{fig10} we plot $(1+G) F$ for different values of the renormalization point $\mu$; 
clearly the dependence on $\mu$ is very mild. The theoretical origin of this residual $\mu$-dependence can be traced back to the approximations used for the ghost-gluon vertex $\gb_{\nu}$ and the  function $H_{\mu\nu}$ (see beginning of subsection C). 
This approximation distorts the multiplicative renormalizability of the corresponding SDEs; indeed, for multiplicative renormalizability to be enforced, 
one must assume the  exact renormalization properties for $\gb_{\nu}$ and $H_{\mu\nu}$, as was done 
in subsection B, where the renormalization was carried out formally. 
Instead, the approximation employed causes a mismatch in higher orders, which introduces the 
observed mild dependence on $\mu$. This dependence can be eliminated by resorting to the systematic improvement of the corresponding  
Ansatz used for $\gb_{\nu}$, in the spirit of the prototype QED calculations presented in \cite{King:1982mk}, and more recently in \cite{Kizilersu:2009kg}.

\subsection{The effective charges}

\begin{figure}[!t]
\begin{center}
\includegraphics[scale=0.8]{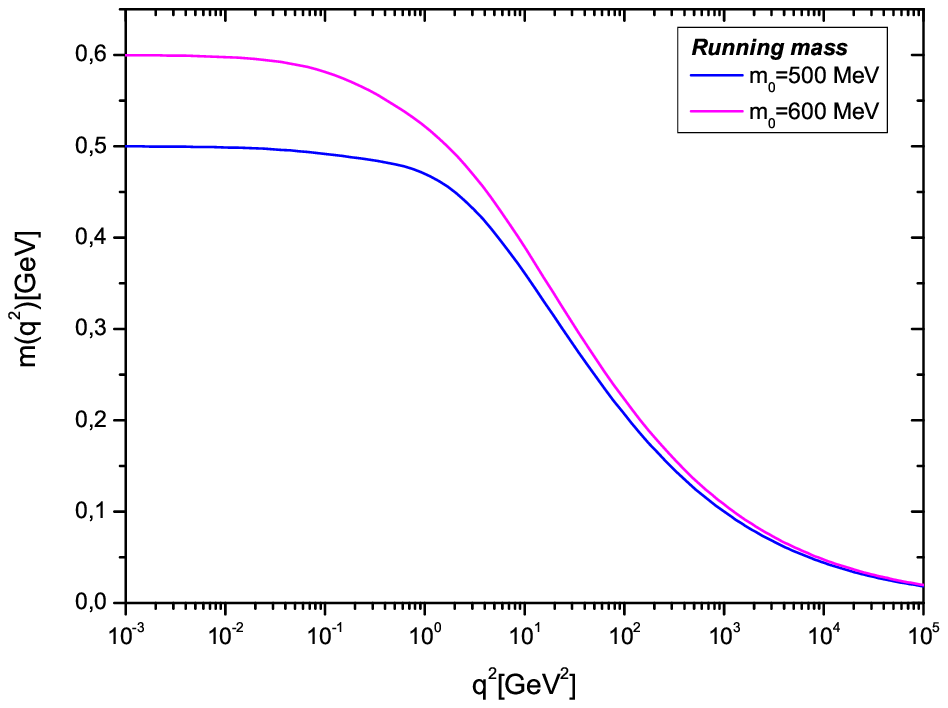}
\end{center}
\caption{The behavior of the running mass given by Eq.~(\ref{rmass}) when \mbox{$m_0=500 \,\mbox{MeV}$} (blue line) 
and \mbox{$m_0=600 \,\mbox{MeV}$} (magenta line). In both cases we used
$\Lambda_\chic{\mathrm{QCD}}=300 \,\mbox{MeV}$.}
\label{fig6}
\end{figure}

We can next proceed to extract the non-perturbative running charge $\alpha_{\chic{\mathrm{PT}}}(q^2)$, defined in Eq.~(\ref{charge}), 
by multiplying the results obtained for $\widehat{d}(q^2)$ by the factor $[q^2 + m^2(q^2)]$. 
To this end, we will assume that $m^2(q^2)$  has a power-law type of running, given by~\cite{Aguilar:2007ie, Lavelle:1991ve}
\be
m^2(q^2)= \frac{m^4_0}{q^2+m^2_0}\left[\ln\left(\frac{q^2 + 2m_0^2}{\Lambda_\chic{\mathrm{QCD}}^2}\right)
\bigg/\ln\left(\frac{2m_0^2}{\Lambda_\chic{\mathrm{QCD}}^2}\right)\right]^3.
\label{rmass}
\ee
Notice that when $q^2\to0$ one has $m^2(0)=m^2_0$. A variety of theoretical and 
phenomenological estimates place it in the range  \mbox{$m_0=350-700 \,\mbox{MeV}$}~\cite{Cornwall:1982zr, Cornwall:2009ud, Bernard:1981pg, Halzen:1992vd}. 
In Fig.~\ref{fig6} we plot the behavior of $m^2(q^2)$ as given by Eq.~(\ref{rmass}), 
for the two values  \mbox{$m_0=500 \,\mbox{MeV}$} and \mbox{$m_0=600 \,\mbox{MeV}$},  
which will be used in the rest of this section.

\begin{figure}[!t]
\begin{minipage}[b]{0.5\linewidth}
\centering
\hspace{-1cm}
\includegraphics[scale=0.8]{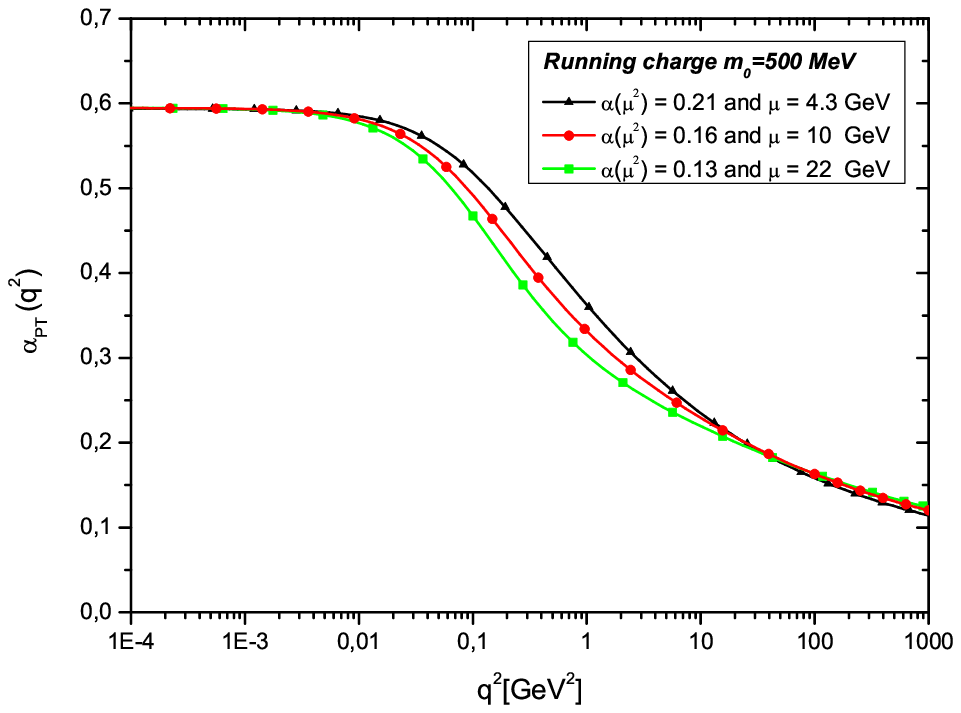}
\end{minipage}
\hspace{0.5cm}
\begin{minipage}[b]{0.45\linewidth}
\centering
\hspace{-2.0cm}
\includegraphics[scale=0.8]{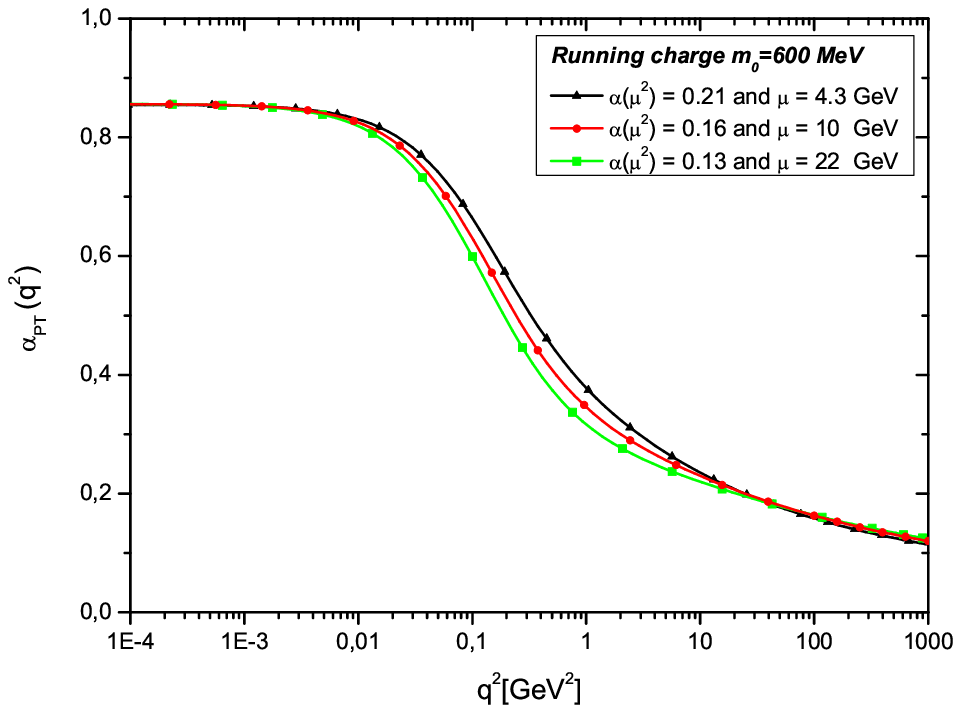}
\end{minipage}
\caption{{\it Left panel}: The running charge obtained from (\ref{charge}) using the SDE solutions
for $\Delta(q^2)$, $D(q^2)$, and $1+G(q^2)$. We use a running mass given
by Eq.~(\ref{rmass}) with \mbox{$m_0=500 \,\mbox{MeV}$}. Right panel: The same for
\mbox{$m_0=600 \,\mbox{MeV}$}.}
\label{fig5}
\end{figure}

On the left panel of  Fig.~\ref{fig5}, we show the results for  
$\alpha_{\chic{\mathrm{PT}}}(q^2)$ when
\mbox{$m_0=500 \,\mbox{MeV}$} in Eq.~(\ref{rmass}).  
The small discrepancy between the three curves is mainly due to the propagation of the tiny residual $\mu$ dependence displayed by the quantity $\widehat d(q^2)$ as shown in Fig.~\ref{fig4}.
One clearly sees that the effective coupling $\alpha_{\chic{\mathrm{PT}}}(q^2)$ freezes out and acquires a finite value in the IR, 
while in the UV it shows the expected  perturbative behavior. For \mbox{$m_0=500 \,\mbox{MeV}$}, 
one gets $\alpha_{\chic{\mathrm{PT}}}(0)\approx 0.6$. One should also notice  that the choice of smaller values of $m_0$ would not produce 
a monotonically decreasing $\alpha_{\chic{\mathrm{PT}}}(q^2)$; instead, one observes the appearance of ``bumps'' in the IR region. 
Therefore if one were to introduce the monotonic 
decrease as an additional requirement of $\alpha_{\chic{\mathrm{PT}}}(q^2)$, 
this would provide a lower bound for the possible values of $m_0$.  
Finally, on the right panel of Fig.~\ref{fig5}, we  show the effective coupling for the case \mbox{$m_0=600 \,\mbox{MeV}$}. 
Now, the freezing occurs at the slightly higher value of $\alpha_{\chic{\mathrm{PT}}}(0)\approx 0.85$. 
Evidently, the freezing value $\alpha_{\chic{\mathrm{PT}}}(0)$ increases as one goes to higher values of $m_0$.  

\begin{figure}[!t]
\begin{minipage}[b]{0.5\linewidth}
\centering
\hspace{-1cm}
\includegraphics[scale=0.8]{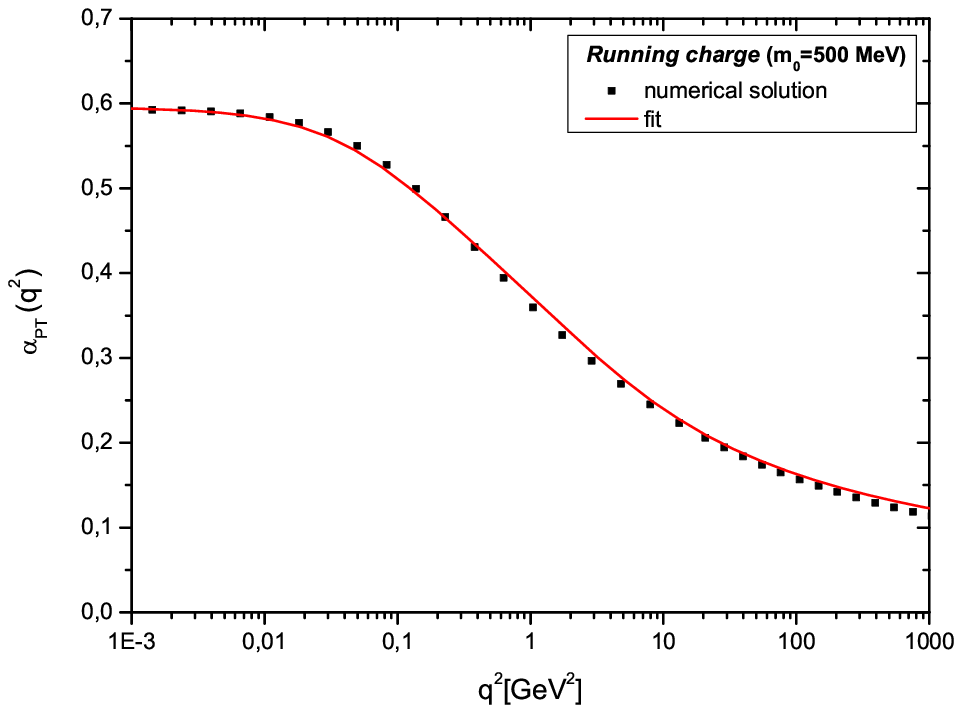}
\end{minipage}
\hspace{0.5cm}
\begin{minipage}[b]{0.45\linewidth}
\centering
\hspace{-2.0cm}
\includegraphics[scale=0.8]{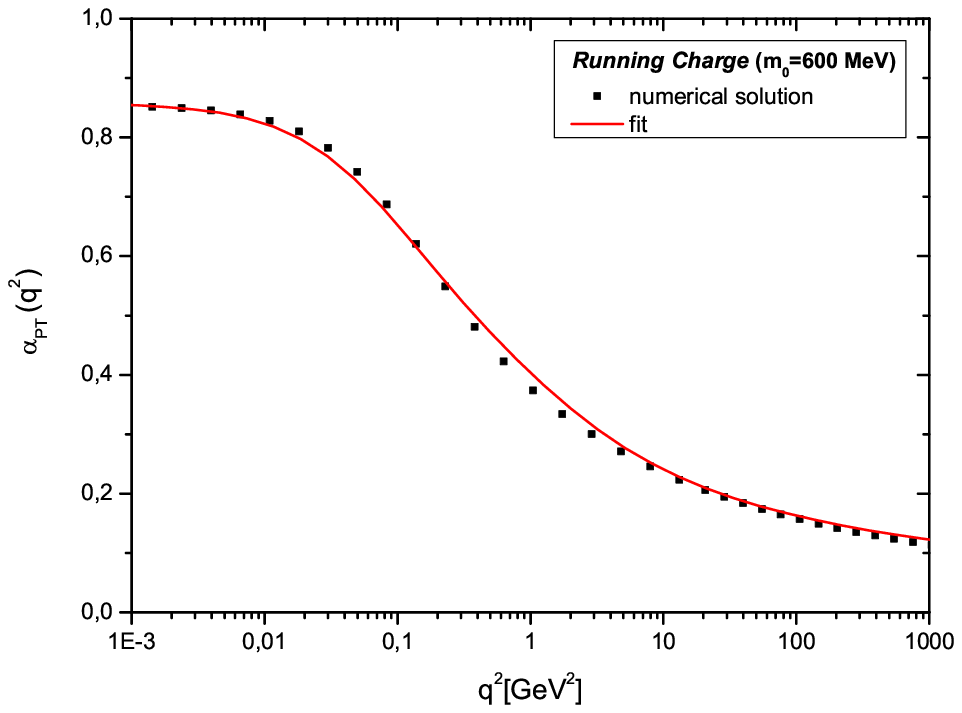}
\end{minipage}
\caption{{\it Left panel}: The fit given by Eq.~({\ref{fit}}) for $m_0=500 \,\mbox{MeV}$; in this case the 
best fit values correspond to $\rho_1=4.5$, and  $\rho_2=-2$. {\it Right panel}: Same as before in the case $m_0=600 \,\mbox{MeV}$; 
in this case the best fit parameters are $\rho_1=2.2$, and  $\rho_2=-1.25$.}
\label{fig7}
\end{figure}

An accurate fit for the running charges shown in Fig.~\ref{fig5} is provided by the following functional form
\be
\alpha(q^2)= \left[4\pi b \ln\left
(\frac{q^2 +h(q^2,m^2(q^2)) }{\Lambda_\chic{\mathrm{QCD}}^2} \right)\right]^{-1},
\label{fit}
\ee     
with the function $h(q^2,m^2(q^2))$ given by
\be
h(q^2,m^2(q^2))= \rho_1m^2(q^2) +\rho_2\frac{m^4(q^2)}{q^2 +m^2(q^2)}.
\label{exfunc}
\ee      
Our best fits to the numerical results for $\alpha_{\chic{\mathrm{PT}}}(q^2)$ using Eq.~(\ref{fit}) above are shown in Fig.~\ref{fig7}.

\begin{figure}[!t]
\begin{minipage}[b]{0.5\linewidth}
\centering
\hspace{-1cm}
\includegraphics[scale=0.8]{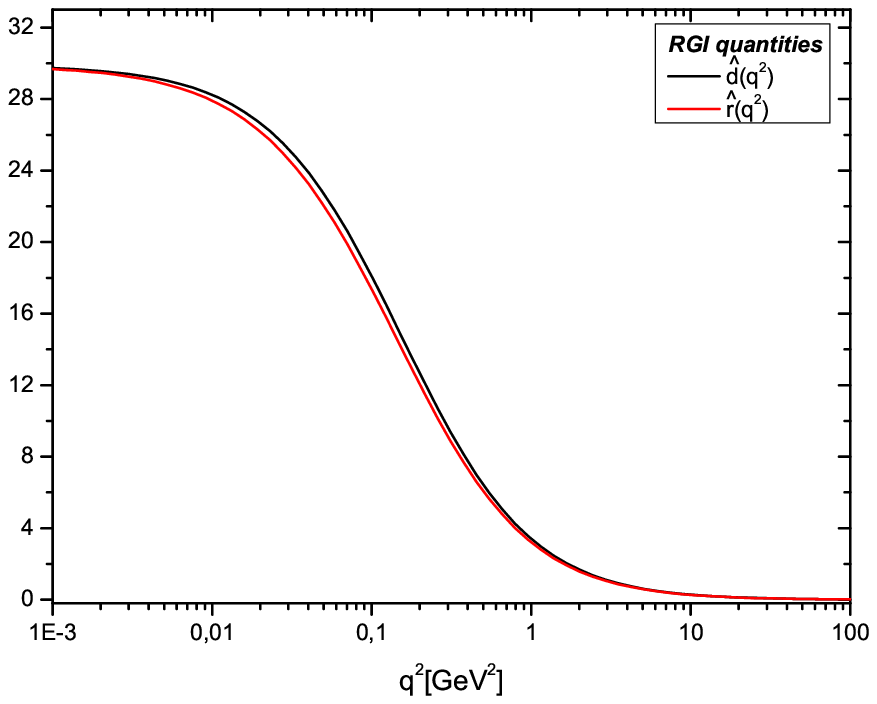}
\end{minipage}
\hspace{0.5cm}
\begin{minipage}[b]{0.45\linewidth}
\centering
\hspace{-2.0cm}
\includegraphics[scale=0.8]{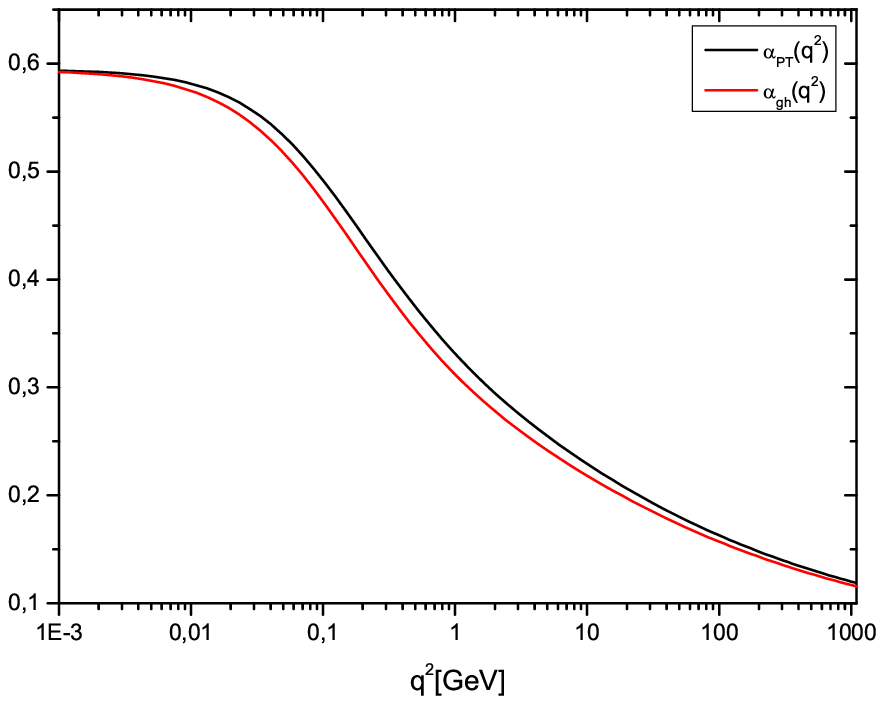}
\end{minipage}
\caption{{\it Left panel}: Comparison between the RG-invariant products $r(q^2)$ (red curve) and  $\widehat d(q^2)$ (black curve). 
{\it Right panel}: Comparison between $\alpha_{\mathrm{gh}}(q^2)$ (red curve) and  $\alpha_{\mathrm{PT}}(q^2)$ (black curve) 
when a mass of  \mbox{$m_0=500 \,\mbox{MeV}$} is factored out. In both plots the difference between the curves appear 
in the intermediate regime of momenta being entirely due to the function $L(q^2)$.}
\label{fig8}
\end{figure}

Finally, we compare numerically the two effective charges, $\alpha_{\chic{\mathrm{PT}}}(q^2)$ and  $\alpha_{\mathrm{gh}}(q^2)$.
The results are shown in Fig.~\ref{fig8},  where $\widehat{r}(q^2)$ is compared with $\widehat{d}(q^2)$ (left panel), 
and $\alpha_{\mathrm{gh}}(q^2)$ with $\alpha_{\chic{\mathrm{PT}}}(q^2)$ (right panel).
As anticipated, the curves coincide in the deep IR and UV, and differ only slightly in the intermediate region. 
To produce both curves, we have factored out a mass of  \mbox{$m_0=500 \,\mbox{MeV}$}, 
whose  dynamical running is again given in Eq.~(\ref{rmass}); equivalently, one could use directly Eq.~(\ref{coup_charge}).

\section{\label{concl}Conclusions}
 
In this article we have presented a detailed comparison between
the two QCD effective charges, $\alpha_\chic{{\mathrm{PT}}}(q^2)$ and $\alpha_{\mathrm{gh}}(q^2)$, 
 obtained within two vastly different 
frameworks: the PT (and BFM) on the one hand, 
and the ghost-gluon vertex (with the Taylor-kinematics) on the other.   
It turns out that their dynamics involves the gluon propagator $\Delta(q^2)$ (in the Landau gauge) 
as a common ingredient, entering in both  
$\alpha_\chic{\mathrm{PT}}(q^2)$ and $\alpha_{\mathrm{gh}}(q^2)$, 
and two different ingredients, 
which participate in a non-trivial identity. 
This identity, which is valid only in the Landau gauge, 
relates the ghost dressing function, $F(q^2)$, with the 
two form-factors,  $G(q^2)$ and $L(q^2)$, 
appearing in the Lorentz decomposition of a special Green's function,  
originating from the ghost sector of the theory.

The two QCD effective charges have been computed 
using as input the non-perturbative solutions of 
a system of three coupled non-linear integral equations, first derived in~\cite{Aguilar:2008xm}, 
containing $\Delta(q^2)$, $F(q^2)$, and $G(q^2)$ as unknown quantities.
The solutions obtained from the 
above system of SDEs for $\Delta(q^2)$ and $F(q^2)$ -- and subsequently fed into the 
defining equations of the effective charges-- 
are in qualitative agreement with recent results from 
large-volume lattices, both for $SU(2)$~\cite{Cucchieri:2007md} and $SU(3)$~\cite{Bogolubsky:2007ud}: 
both quantities reach finite (non-vanishing) values in the deep IR.    
One important consequence of the central identity (and the dynamics encoded in the relevant equations) 
is that the two charges are 
identical not only in the deep UV, where asymptotic freedom manifests itself, 
but also in the deep IR, where they ``freeze'' at the same non-vanishing value.

As already mentioned in section \ref{numan}, 
at the level of the SDE for the gluon propagator, namely Eq.~(\ref{sdef}), 
the value of 
$\Delta (0)$ is a free parameter. The value chosen for $\Delta (0)$ affects (in a non-linear way)   
the IR values of the RG-invariant quantities, namely ${\widehat d}(0)$ and ${\widehat r}(0)$, 
which, in turn, restricts the values of the gluon mass, $m_0$, and the 
freezing value of the effective charges. Throughout the analysis presented in section \ref{numan} 
the criterion used for choosing the values of $\Delta (0)$ was that 
the resulting values for $m_0$ and 
$\alpha_{\mathrm{gh}}(0)$, (or $\alpha_\chic{\mathrm{PT}}(0)$) would be numerically compatible 
with those obtained from a variety of phenomenological studies~\cite{Halzen:1992vd}.  
Specifically,  values for $m_0$ in the range of $350-700\,\mbox{MeV}$ and 
$\alpha_\chic{\mathrm{PT}}(0) \approx 0.7\pm 0.3$. 
Notice, however, a subtle point that may be of relevance when carrying out such comparisons.
The effective charge assumed in most of the aforementioned studies 
is that of~\cite{Cornwall:1982zr}, which has a very particular functional form, 
and corresponds to the standard PT construction, 
where the Feynman gauge of the BFM is dynamically singled out. Instead, for the reason explained 
in subsection 2.2, the present analysis is based on the generalized PT~\cite{Pilaftsis:1996fh}, which 
eventually projects one to the Landau gauge of the BFM. It would be interesting 
to reach a quantitative understanding of how the aforementioned difference in the gauges 
affects the phenomenological 
values of the gluon mass and of the freezing of the effective charge.  
Calculations in this direction are already in progress.

As has been emphasized in~\cite{Aguilar:2008xm}, 
even though the solutions of the SDE system 
are in qualitative agreement with the aforementioned lattice results, 
they display a considerable quantitative discrepancy from them.
Specifically, $\Delta(q^2)$ differs significantly in the region of intermediate momenta,   
and the value of the ghost dressing function is about a factor 
of two less than that obtained on the lattice. 
These discrepancies, in turn, are expected to affect the numerical values 
(but {\it not} the qualitative features) of quantities  
computed using them as input. In particular, it should be interesting to 
obtain the QCD effective charges studied here using as input 
the lattice results for $\Delta(q^2)$ and $F(q^2)$, and 
[indirectly, using, {\it e.g.}, the first equation in (\ref{LFGr})] for $G(q^2)$; 
we hope to address this issue in a future work.   

\acknowledgments
 
The research of J.~P. is supported by the European FEDER and  Spanish MICINN under grant FPA2008-02878, and the Fundaci\'on General of the UV. The work of J. R-Q. is supported  by the Spanish MEC grant FPA-2006-13825.

\end{document}